\documentclass[aps,prc,letterpaper,11pt,twoside,tightenlines,nofootinbib,showpacs,preprint,superscriptaddress]{revtex4}
\usepackage{graphicx}
\usepackage[sort&compress]{natbib}
\usepackage{subfigure}
\usepackage{amsmath}
\usepackage{amsfonts}
\usepackage{cancel}

\newcommand{\Ima}{\textrm{Im}}
\newcommand{\mev}{\textrm{ MeV}}
\newcommand{\gev}{\textrm{ GeV}}
\newcommand{\rb}[1]{\raisebox{2.2ex}[0pt]{#1}}

\begin{document}

\title{States generated in the $K$-multi-$\rho$ interactions}

\author{C. W. Xiao}
\affiliation{Institute for Advanced Simulation and Institut  f\"{u}r Kernphysik (Theorie), Forschungszentrum J\"ulich, D-52425 J\"ulich, Germany}
\affiliation{Departamento de F\'{\i}sica Te\'orica and IFIC, Centro Mixto Universidad \\de Valencia-CSIC, Institutos de Investigaci\'on de Paterna, Apartado 22085, 46071 Valencia, Spain}

\date{\today}

\begin{abstract}

In the present work we use three-body interaction formalism to investigate the $K$-multi-$\rho$ interactions. First, we reproduced the resonances $f_2(1270)$ and $K_1(1270)$ in the $\rho \rho$ and $\rho K$ two-body interactions respectively, as the clusters of the fixed center approximation. Then, we study the three-body $K -\rho \rho (f_2)$ and $\rho - \rho K (K_1)$ interactions with the fixed center approximation of the Faddeev equations. Furthermore, we extrapolate the formalism to study the four-body, five-body and six-body systems, containing one $K$ meson and multi-$\rho$ mesons. In our research, without introducing any free parameters, we generate the $K_2 (1770)$ state in the three-body interaction with the mass of $1707\mev$ and a width about $113\mev$, which are consistent with the experiments. We also find a clear resonant structure in our results of the five-body interaction, with a mass $2505\mev$ and a width about $32\mev$ or more, which is associated to the $K_4 (2500)$ state,  where we obtain consistent results with the experimental findings. Furthermore, we predict some new states in the other many body interactions, $K_3 (2080)$, $K_5 (2670)$ (isospin $I=1/2$), and $K_4 (2640)$ (isospin $I=3/2$), with uncertainties.

\end{abstract}


\maketitle

\section{Introduction}

To understand the nature and structure of the particles found in the experiments, and to search for new hadronic states, are the main issues in nowadays particle physics which has been accepted generally that quarks are the basic building blocks of matter. Within the Gell-Mann-Zweig quark model \cite{Feynman:1964fk,GellMann:1964nj} for the normal hadron states, mesons are made of a quark-antiquark pair, $q \bar{q}$, and baryons made of three quark components, $qqq$. On the other hand,  there are some states found in the experiments, such as the mesons, $f_0 (500)$, $f_0 (980)$, $a_0 (980)$, $\kappa (800)$, the baryons, $\Lambda (1405)$, $N (1440)$, $N (1535)$, with structure and properties difficult to explain by the normal quark model. These states might therefore be called ``exotic'' states (recent experimental discussions are given in Refs. \cite{Olsen:2014mea,Choi:2014iwa}). To understand the structure and properties of these exotic states in the strong interaction, we need to exploit other theories or approaches, for example chiral perturbative theory \cite{Gasser:1984gg,Meissner:1993ah,Bernard:1995dp,Pich:1995bw,Ecker:1994gg,Scherer:2002tk,Bernard:2007zu}, effective field theory \cite{Politzer:1988bs,Georgi:1990um,Epelbaum:2008ga}, Lattice QCD \cite{Kogut:1982ds,Luscher:1996sc,Luscher:1996ug}, QCD sum rule \cite{Shifman:1978bx,Reinders:1984sr,Dias:2012ek,Zhou:2014ytp}, Dyson-Schwinger equations \cite{Roberts:1994dr,Maris:2003vk,Fischer:2006ub}, chiral quark model \cite{Manohar:1983md,Zhang:1997ny,Fontoura:2012mz}, chiral unitary approach (ChUA) \cite{Oller:1997ti,Oset:1997it,Oller:1997ng,Oller:1998hw,Oller:1998zr,Oller:2000fj}, and so on. But, some discovered particles, such as the $\phi (2170)$ (also called $X (2175)$, $Y (2175)$), $Y (4260)$, $N^* (1710)$, look like having a more complicated structure and could come from multi-body hadron interaction, which is a subject in hadron physics drawing much attention for a long time \cite{Fujita:1957zz,Faddeev:1960su,Glockle:1986zz,Weinberg:1992yk,Richard:1992uk}. With this motivation, the work of Ref. \cite{MartinezTorres:2007sr} develops the ChUA for the three-body interaction, which combines the three-body Faddeev equations with on shell approximation of the ChUA and has reported several $S$-wave $J^P=\frac{1}{2}^+$ resonances qualifying as two mesons-one baryon molecular states. In Ref. \cite{MartinezTorres:2012jr} this combination of Faddeev equations and chiral dynamics in the $D K \bar K$ system obtains consistent results with QCD sum rules. When in some cases there are resonances (or bound states) appearing in the two-body subsystem of the three-body interaction, Ref. \cite{Roca:2010tf} takes the fixed center approximation (FCA) \cite{Faddeev:1960su,Toker:1981zh,Barrett:1999cw,Deloff:1999gc,Kamalov:2000iy,Gal:2006cw} to the Faddeev equations, where several multi-$\rho(770)$ states are dynamically produced, and the resonances $f_2(1270)(2^{++})$, $\rho_3(1690)(3^{--})$, $f_4(2050)(4^{++})$, $\rho_5(2350)(5^{--})$, and $f_6(2510)(6^{++})$ are theoretically found as basically molecules of an increasing number of $\rho(770)$ particles with parallel spins. Analogously, the resonances $K^*_2(1430)$, $K^*_3(1780)$, $K^*_4(2045)$, $K^*_5(2380)$ and a new $K^*_6$ are produced in the $K^*$-multi-$\rho$ systems and could be explained as molecules with the components of an increasing number of $\rho(770)$ and one $K^*(892)$ meson in Ref. \cite{YamagataSekihara:2010qk}. Also in Ref. \cite{Xiao:2012dw}, charmed resonances $D^*_3$, $D^*_4$, $D^*_5$ and $D^*_6$ are predicted in the $D^*$-multi-$\rho$ interaction. Note that, recently a resonance structure is found by LHCb at about $2.8 \gev$ with $J^P=3^-$ in the $\bar{D}^0 \pi^-$ mass distribution, which is close to the mass of $D^*_3$ state predicted in Ref. \cite{Xiao:2012dw}, $2800 \sim 2850 \mev$. In the present work we investigate the $K$-multi-$\rho$ interaction.
    
Taking FCA to the Faddeev equations, the $\bar{K} NN$ interaction studied in Refs. \cite{Bayar:2011qj,Bayar:2012rk} has proved accuracy when dealing with bound states, and obtains consistent results with the full Faddeev equations evaluation without taking FCA \cite{MartinezTorres:2008kh}, or a variational calculation with a nonrelativistic three-body potential model \cite{Jido:2008kp}, which is also confirmed by the recent results with new Faddeev calculations in Refs. \cite{Shevchenko:2014uva,Revai:2014twa}. Even though there is a different results claimed in Ref. \cite{Miyagawa:2012xz} on the $\bar{K} NN$ system, the work of Ref. \cite{Jido:2012cy} clarified the different kinematical between them using two different approaches in their investigations, the Watson approach and the truncated Faddeev approach. A further study of the $\bar{K} NN$ system is done in the recent work of  \cite{Mai:2014uma}, which has investigated the $\bar{K} d$ scattering length with the first-order recoil correction using the non-relativistic effective field theory approach.  A narrow quasibound state of $3500 \mev$ in the $DNN$ system is predicted in Ref. \cite{Bayar:2012dd} by both FCA to Faddeev equations calculation and the variational method approach with the effective one-channel Hamiltonian. Therefore, in the present work, we use the FCA to the Faddeev equations to investigate the $K$-multi-$\rho$ interaction. There are some possible $K$ excited states with strangeness $S=\pm 1$ and aligned with increasing spin number in the Particle Data Group (PDG) \cite{pdg2014}, such as $K_1 (1270) (1^+)$ or $K_1 (1400) (1^+)$ or $K_1 (1650) (1^+)$,  $K_2 (1580) (2^-)$ or $K_2 (1770) (2^-)$ or $K_2 (1820) (2^-)$ or $K_2 (2250) (2^-)$, $K_3 (2320) (3^+)$, $K_4 (2500) (4^-)$. Some of these states still need more confirmation in the future experiments. This is the motivation of the present work, to understand their structure and properties theoretically. Applying the FCA to the Faddeev equations, there should be resonances or bound states in the two-body subsystem. For the $K$-multi-$\rho$ systems, the basic two-body subsystems are the $\rho \rho$ and $K \rho$ interactions. Based on the local hidden gauge Lagrangians \cite{Bando:1984ej,Bando:1987br,Meissner:1987ge,Harada:2003jx}, the two-body $\rho\rho$ interaction is studied in Ref. \cite{Molina:2008jw} with the coupled channel approach of ChUA and found that a $\rho\rho$ quasibound state or molecule could be associated to the $f_2(1270)$ found in the PDG \cite{pdg2014}. With the on-shell Bethe-Salpeter equation of ChUA, and a chiral Lagrangian, the two-body $K \rho$ interaction is studied in Ref. \cite{Roca:2005nm} and dynamically produced the $K_1 (1270)$ resonance. Furthermore, the $K_1 (1270)$ state is reinvestigated with detail using ChUA to analyse the experimental data in the later work of Ref. \cite{Geng:2006yb}. Thus, the resonances $f_2(1270)$ in the $\rho\rho$ interaction and $K_1 (1270)$ in the $K \rho$ interaction are the needed clusters in the formalism of the present work.

In the next section, we will first present the formalism of the FCA to the Faddeev equations. Then, in the following section, the resonances $f_2(1270)$ and $K_1 (1270)$ are reproduced dynamically with the ChUA in the $\rho\rho$ interaction and the $K \rho$ interaction respectively. Our investigation results are shown in Section \ref{secres}. Finally, we finish with conclusions.

\section{Formalism}
\label{secform}

\begin{figure}
\centering
\includegraphics[scale=0.3]{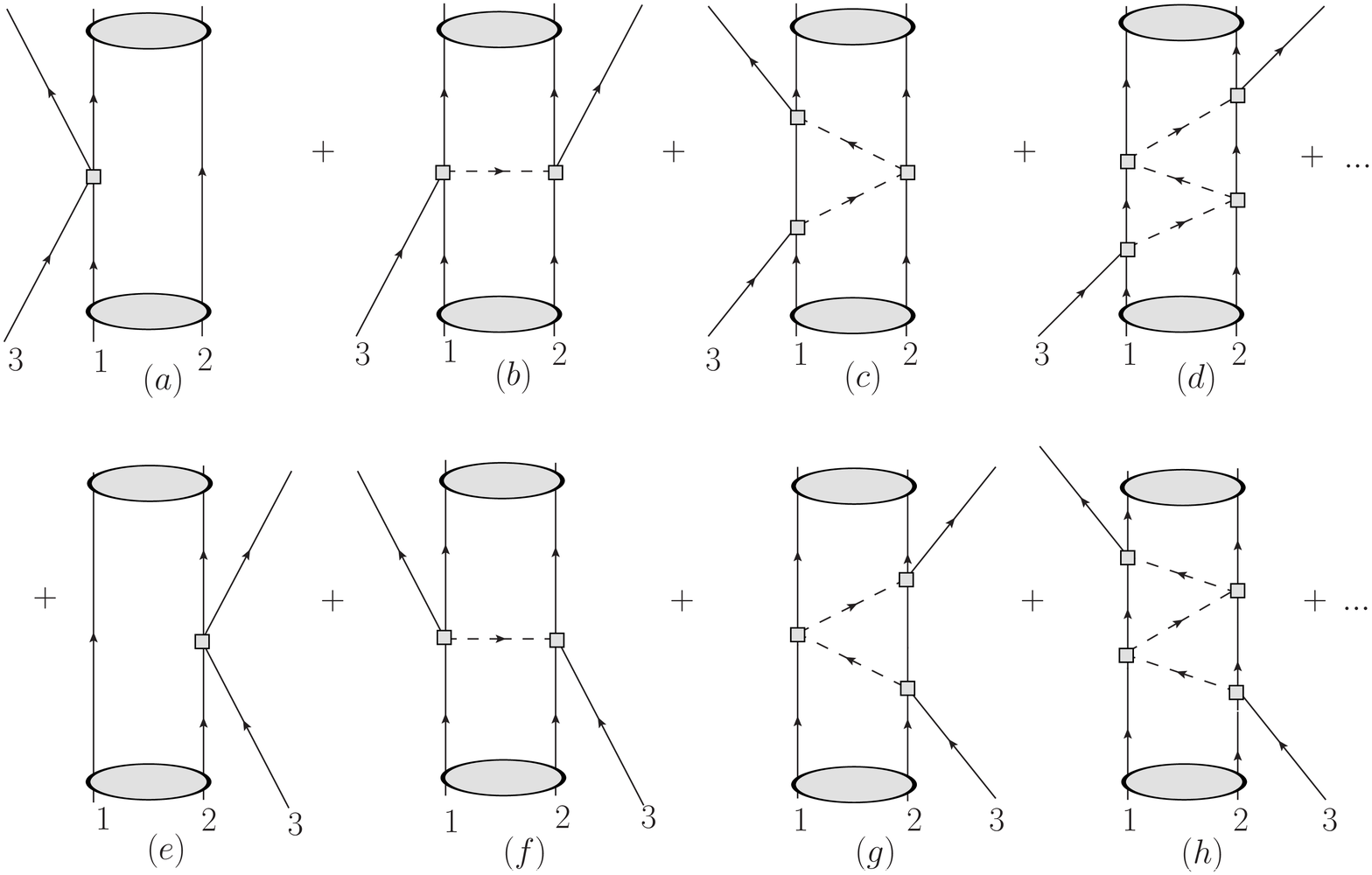}
\caption{Diagrammatic representation of the FCA to Faddeev equations.}\label{fig:FCA}
\end{figure}

For the three-body interaction as Faddeev suggested in Ref. \cite{Faddeev:1960su}, the scattering amplitude of $T$-matrix can be written as a sum of three partitions,
\begin{equation}
T=T^1+T^2+T^3,
\label{eq:faddeev}
\end{equation}
where partition amplitude $T^i$ ($i=1$, $2$, $3$) includes all the possible interactions contributing to the three-body $T$-matrix with the particle $i$ being a spectator in the last interaction. But, if there are resonances (or bound states) as clusters appearing in the two-body subsystem interaction, for example the cluster coming from $T^3$, thus, we can assume that a cluster is formed by the two particles (named as particle 1, 2) and is not much modified by the interaction of a third particle (particle 3) with this cluster. Therefore, assuming the cluster as the fixed center of the three-body system, we can take the FCA \cite{Faddeev:1960su,Toker:1981zh,Barrett:1999cw,Deloff:1999gc,Kamalov:2000iy,Gal:2006cw} to the Faddeev equations. Then, the $T^3$ partition amplitude contributes to the cluster for the FCA, and the FCA multiple scattering of the third particle with the components of the cluster is taken into account. Thus, we can rewrite the Faddeev equations of Eq. \eqref{eq:faddeev} easily (which is also developed by the ChUA),
\begin{align}
T_1&=t_1+t_1G_0T_2,\\
T_2&=t_2+t_2G_0T_1,\\
T&=T_1+T_2,
\end{align}
where $T$ is the total three-body scattering amplitude, as depicted in Fig. \ref{fig:FCA}.
From this figure, we can see that the Faddeev equations under the FCA are first a pair of particles (1 and 2) forming a cluster, and then particle 3 interacts with the components of the cluster, undergoing all possible multiple scattering with those components. Thus, the two partition amplitudes $T_1$ and $T_2$ sum all diagrams of the series of Fig. \ref{fig:FCA} which begin with the interaction of particle 3 with particle 1 of the cluster ($T_1$), or with the particle 2 ($T_2$). Finally, the scattering amplitude $T$ is the total three-body interaction amplitude that we look for. The amplitudes $t_1$ and $t_2$ represent the unitary scattering amplitudes with coupled channels for the interactions of particle 3 with particle 1 and 2, respectively, which should be taken into account the isospin structure of the subsystem and discussed in details for different cases in Section \ref{secres}. Besides, $G_0$ is the propagator of particle 3 between the components of the two-body subsystem, given by
\begin{equation}
G_0(s)= \frac{1}{2 M_R} \int \frac{d^3\vec{q}}{(2\pi)^3} F_R(\vec{q}\,) \frac{1}{q^{02}(s)-\vec{q}^{~2}-m_3^2 +i\,\epsilon},
\label{eq:G0s}
\end{equation}
where $q^0 (s)$, the energy carried by particle 3 in the rest frame of the three particle system, is given by
\begin{equation}
q^0(s)=\frac{s+m_3^2-M_R^2}{2\sqrt{s}},
\end{equation}
with $m_3$ the mass of the third particle and $M_R$ the mass of the cluster, and $F_R(\vec{q}\,)$ is the form factor of the cluster of particles 1 and 2. The form factor of the cluster should be consistently with the theory used to generate the cluster. This requires to evaluate the wave functions in the ChUA reproduced the cluster, which has been done in Refs. \cite{Gamermann:2009uq,YamagataSekihara:2010pj,Aceti:2012dd} for $S$-wave bound states, $S$-wave resonant states and states with arbitrary angular momentum, respectively. Since in the present cases, the generated clusters are $S$-wave bound states, we only need the expression of the form factors for the $S$-wave bound states, given by \cite{YamagataSekihara:2010pj},
\begin{align}
\begin{split}
F_R(\vec{q}\,)&=\frac{1}{\mathcal{N}} \int_{|\vec{p}\,|<\Lambda', |\vec{p}-\vec{q}\,|<\Lambda'} d^3 \vec{p} \; \frac{1}{2 \omega_1(\vec{p}\,)} \frac{1}{2 \omega_2(\vec{p}\,)} \frac{1}{M_R - \omega_1(\vec{p}\,) -\omega_2(\vec{p}\,)} \\
&\quad\frac{1}{2 \omega_1(\vec{p}-\vec{q}\,)} \frac{1}{2 \omega_2(\vec{p}-\vec{q}\,)} \frac{1}{M_R-\omega_1(\vec{p}-\vec{q}\,)-\omega_2(\vec{p}-\vec{q}\,)}, \label{eq:formfactor}
\end{split}\\
\mathcal{N}&=\int_{|\vec{p}\,|<\Lambda'} d^3 \vec{p} \; \Big( \frac{1}{2 \omega_1(\vec{p}\,)} \frac{1}{2 \omega_2(\vec{p}\,)} \frac{1}{M_R-\omega_1(\vec{p}\,)- \omega_2(\vec{p}\,)} \Big)^2, \label{eq:formfactorN}
\end{align}
where $\omega_i = \sqrt{\vec{q}\,^2 + m_i^2}$ ($i=1, \, 2$, $m_i$ the mass) are the energies of the particles 1, 2. We use a cut off $\Lambda'$ to regularize the integrals of Eqs. \eqref{eq:formfactor} and \eqref{eq:formfactorN} (also for Eq. \eqref{eq:G0s}, discussed in Section \ref{secres}), which is the same as the one used in the loop function of the two-body interaction to reproduce the cluster \cite{YamagataSekihara:2010qk,Xie:2010ig}. Thus, no free parameters are involved. Note that, in the present work, we care about the dynamics close to the threshold and thus the method of cutoff is acceptable  \footnote{We use a common cut-off method to regularize the propagator. Since it violates gauge invariance, it is a effective way to remove infinities from perturbative calculations, of which the problem is addressed in Ref.~\cite{Polchinski:1983gv} within the effective Lagrangian. Furthermore, respecting Lorentz invariance, the cut-off regularization can recover the symmetry~\cite{Varin:2006de} (references therein).}. 

Taking the normalization of the field theory \cite{mandl} which has different weight factors for the particle fields, we must take into account how these factors appear in the single scattering and double scattering and in the total amplitude \cite{YamagataSekihara:2010qk,Xie:2010ig}. In all the present cases, the cluster (also particles 1 and 2) is a meson and the scattering particle (the third particle) too, which are only related to meson fields. Thus, following Ref. \cite{mandl} we write the $S$ matrix of single scattering, Fig. \ref{fig:FCA} (a),
\begin{align}
S^{(1)}_1=&-it_1 (2\pi)^4\,\delta(k+k_R-k'-k'_R) \nonumber \\
&\times \frac{1}{{\cal V}^2} \frac{1}{\sqrt{2\omega_3}} \frac{1}{\sqrt{2\omega'_3}}
 \frac{1}{\sqrt{2\omega_1}} \frac{1}{\sqrt{2\omega'_1}},\label{eq:s11}\\
S^{(1)}_2=&-it_2 (2\pi)^4\,\delta(k+k_R-k'-k'_R)  \nonumber \\
&\times\frac{1}{{\cal V}^2} \frac{1}{\sqrt{2\omega_3}} \frac{1}{\sqrt{2\omega'_3}}
 \frac{1}{\sqrt{2\omega_2}} \frac{1}{\sqrt{2\omega'_2}},\label{eq:s12}
\end{align}
where, $k,\,k'$ ($k_R,\,k'_R$) are the momentum of initial, final scattering particle ($R$ for the cluster), $\omega_i,\,\omega'_i$ are the energies of the initial, final particles,  $\cal V$ is the volume of the box where the states are normalized to unity and the subscripts 1, 2 refer to scattering with particle 1 or 2 of the cluster.

Next, the double scattering diagram, Fig. \ref{fig:FCA} (b), is given by,
\begin{align}
S^{(2)}=&-i(2\pi)^4 \delta(k+k_R-k'-k'_R) \frac{1}{{\cal V}^2}
\frac{1}{\sqrt{2\omega_3}} \frac{1}{\sqrt{2\omega'_3}}
 \frac{1}{\sqrt{2\omega_1}} \frac{1}{\sqrt{2\omega'_1}}
 \frac{1}{\sqrt{2\omega_2}} \frac{1}{\sqrt{2\omega'_2}} \nonumber \\
&\times\int \frac{d^3q}{(2\pi)^3} F_R(\vec{q}\,) \frac{1}{{q^0}^2-\vec{q}\,^2-m_3^2+i\,\epsilon} t_{1} t_{2},\label{eq:s2}
\end{align}
where $F_R(\vec{q}\,)$ is the cluster form factor that we have discussed above, seen in Eq. \eqref{eq:formfactor}.

Similarly, the full $S$ matrix for scattering of particle 3 with the cluster can be written as,
\begin{equation}
S=-i\, T \, (2\pi)^4 \delta(k+k_R-k'-k'_R)\times\frac{1}{{\cal V}^2}
\frac{1}{\sqrt{2 \omega_3}} \frac{1}{\sqrt{2 \omega'_3}}
\frac{1}{\sqrt{2\omega_R}} \frac{1}{\sqrt{2\omega'_R}}.\label{eq:sful}
\end{equation}
Now, we can see that for the unitary amplitudes corresponding to single-scattering contribution, one must take into account the isospin structure of the cluster and write the $t_1$ and $t_2$ amplitudes in terms of the isospin amplitudes of the (3,1) and (3,2) systems. In view of the different normalization of these terms by comparing Eqs. \eqref{eq:s11}, \eqref{eq:s12}, \eqref{eq:s2} and \eqref{eq:sful}, we can introduce suitable factors in the elementary amplitudes,
\begin{equation}
\tilde{t_1}=\frac{2M_R}{2m_1}~ t_1,~~~~\tilde{t_2}=\frac{2M_R}{2m_2}~ t_2,
\label{eq:t1t2}
\end{equation}
where we have taken the approximations, $\frac{1}{\sqrt{2 \omega_i}} \simeq \frac{1}{\sqrt{2m_i}}$. But, when the cluster, the particles 1 and 2, one of them is a baryon, the factors in Eqs. \eqref{eq:G0s} and \eqref{eq:t1t2}, $2 M_R$ and $2 m_i$ should be replaced by 1 for taking the baryonic field factor approximation $\sqrt{\frac{2M_B}{2E_B}} \approx 1$. Finally, we sum all the diagrams, obtained
\begin{equation}
T=T_1+T_2=\frac{\tilde{t_1}+\tilde{t_2}+2~\tilde{t_1}~\tilde{t_2}~G_0}{1-\tilde{t_1}~\tilde{t_2}~G_0^2}. \label{eq:new}
\end{equation}
When $\tilde{t_1}= \tilde{t_2}$ in some cases, it can be simplified as,
\begin{equation}
T=\frac{2 \, \tilde{t_1}}{1-\tilde{t_1} \, G_0}. \label{eq:new2}
\end{equation}
Note that, the FCA to Faddeev equations are particularly suited to study system with the subsystem bound or even loose bound, as discussion in Refs. \cite{Bayar:2011qj,Bayar:2012rk}, and have some limitation on the case when the cluster of the two particles in the subsystem is excited in the intermediate states (more discussions seen in Ref. \cite{MartinezTorres:2010ax} for the study of $\phi K \bar{K}$ system).

From the $S$ matrix of single scattering, Eqs.  \eqref{eq:s11}, \eqref{eq:s12}, and the full $S$ matrix, Eq. \eqref{eq:sful}, we should note that the arguments of the amplitudes $T_i(s)$ and $t_i(s_i)$ are different, where $s$ is the total invariant mass of the three-body system, and $s_i$ are the invariant masses in the two-body subsystems. The relationship between them is given by \cite{YamagataSekihara:2010qk},
\begin{equation}
s_i=m_3^2+m_i^2+\frac{(M_R^2+m_i^2-m_j^2)(s-m_3^2-M_R^2)}{2M_R^2}, (i,j=1,2,\;i\neq j).
\label{eq:si}
\end{equation}

\section{Two-body interaction}
\label{sectwo}

Using the Faddeev equations under the FCA, we first need bound states in the two-body subsystem as the cluster of the fixed center, and then let the third particle collide with the cluster and interact with the components of the forming cluster. As discussed in the introduction, for the subsystems of the $K$-multi-$\rho$ system, the two-body $\rho\rho$ and $\rho K$ interactions are studied in Refs. \cite{Molina:2008jw} and \cite{Geng:2006yb}. We briefly summarize the works of them to reproduce the resonances $f_2 (1270)$ and $K_1 (1270)$, and obtain the two-body scattering amplitudes of the subsystem (details given in the appendix).

\subsection{$\rho\rho$ interaction}
\label{subsecRR}

In Ref. \cite{Molina:2008jw}, the $\rho\rho$ interaction is studied with the local hidden gauge formalism \cite{Bando:1984ej,Bando:1987br,Meissner:1987ge,Harada:2003jx} and the ChUA with coupled channels, and dynamically produced the $f_2(1270)$ ( $I(J^{PC})=0(2^{++})$ ) state. From the local hidden gauge Lagrangians, the $s$-wave potentials of spin $S=2$ for $\rho\rho$ interaction are obtained in the sectors of isospin $I=0$ and $I=2$,
\begin{eqnarray}
V^{(I=0,S=2)}_{\rho\rho} (s_i) &=& -4 g^2 - 8 g^2 \Big( \frac{3s_i}{4m_\rho^2} - 1 \Big), \\
V^{(I=2,S=2)}_{\rho\rho} (s_i) &=& 2 g^2 + 4 g^2 \Big( \frac{3s_i}{4m_\rho^2} - 1 \Big), 
\end{eqnarray}
where $g=M_V/2f_\pi$, with $M_V$ the vector meson mass and $f_\pi$ the pion decay constant. The scattering amplitude of $\rho\rho$ interaction is calculated by the on-shell Bethe-Salpeter equation,
\begin{equation}
t^I = [1-V^I G^I]^{-1} V^I,
\label{eq:bse}
\end{equation}
where the kernel $V^I$ is a matrix of the interaction potentials and $G^I$ a diagonal matrix of the loop functions (upper index $I$ represents the specific isospin sector).

\begin{figure}
\centering
\includegraphics[width=0.6\textwidth]{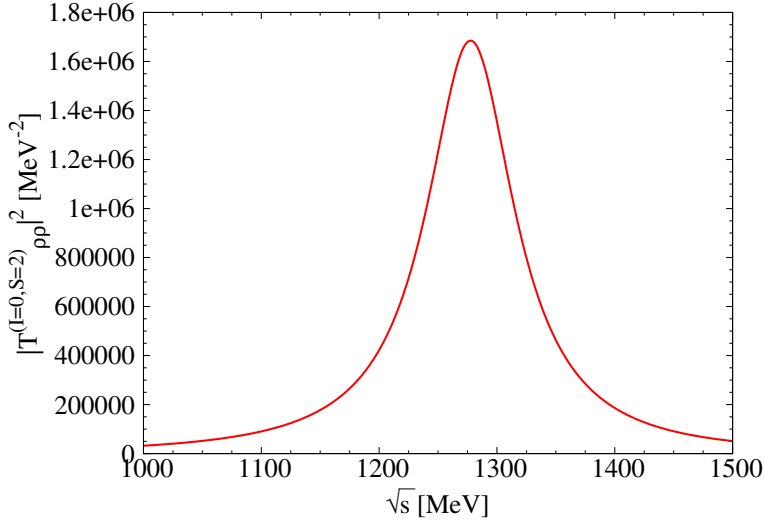}
\caption{Modulus squared of the scattering amplitudes: $|t_{\rho\rho}^{I=0}|^2,\; f_2(1270)$ showing up.}\label{fig:trhorho}
\end{figure}

Following the work of Ref. \cite{Molina:2008jw} (more details seen Appendix \ref{appa}), we also take into account the contribution of the box diagram with two pseudoscalar mesons in the intermediate state. We only consider the imaginary part of the box diagram contribution as the correction of the potential $V^I$, and neglect its real part which is very small. Note that we do not take these intermediate channels in the box diagram (more detail seen in Ref. \cite{Molina:2008jw}) as accounting for the coupled channels \cite{Wu:2010jy,Wu:2010vk}. On the other hand, as done in Ref. \cite{Molina:2008jw}, we also consider the $\rho$ mass distribution by replacing the loop function in the corresponding channel with its convoluted form. We obtain consistent results with Ref. \cite{Molina:2008jw}, as shown in Fig. \ref{fig:trhorho}, where the structure of the resonance $f_2(1270)$ is found in the peak of the modulus squared of the amplitude. We have successfully reproduced the $f_2(1270)$ state as the cluster of our procedure. Besides, the nonresonant amplitude $t_{\rho\rho}^{(I=2,S=2)}$ is not shown in the figure, which is needed when we evaluate the two-body interaction amplitudes $t_1, \, t_2$, considering the isospin structure of the subsystem (discussed in Section \ref{secres}).

\subsection{$\rho K$ interaction}
\label{subsecRK}

\begin{figure}
\centering
\includegraphics[width=0.6\textwidth]{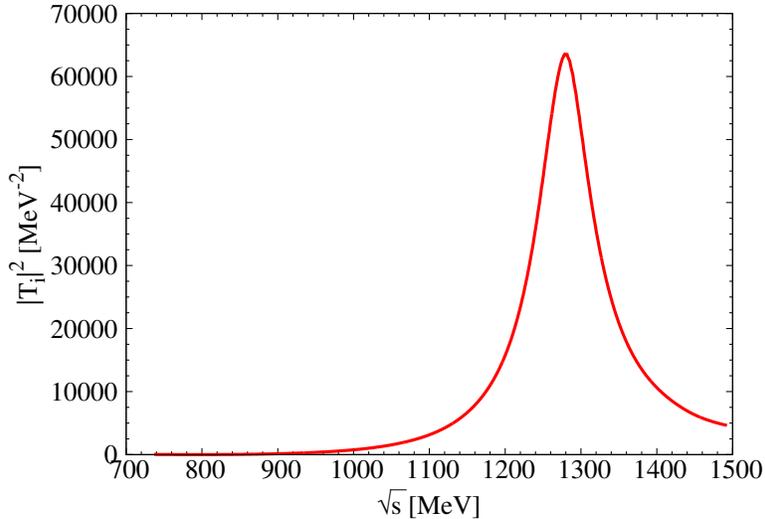}
\caption{Modulus squared of the scattering amplitudes: $|t_{\rho K}^{I=1/2}|^2,\; K_1(1270)$ showing up.}\label{fig:trhoK}
\end{figure}

The $\rho K$ interaction is investigated by the ChUA in Refs. \cite{Roca:2005nm,Geng:2006yb}. Following the work of Ref. \cite{Geng:2006yb}, we can dynamically generate the $K_1 (1270)$ resonance in the interaction of $\rho K$ and its couple channels, $\phi K$, $\omega K$, $K^* \eta$ and $K^* \pi$. From the local hidden gauge Lagrangians, the vector-pseudoscalar potential projected over $s$-wave can be written as
\begin{eqnarray}
  V_{ij}(s_i)  &=&
  -\frac{\vec{\epsilon}\cdot \vec{\epsilon}\;'}{8 f^2} C_{ij}
  \left[ 3s_i - (M_i^2 + m_i^2 + M_j^2 + m_j^2)
  \right. \nonumber \\
  &&
  \left.
  - \frac{1}{s_i}(M_i^2 - m_i^2)(M_j^2 - m_j^2) \right] ,
\label{eq:V_VVPP} 
\end{eqnarray}
where $M_{i(j)}$ and $m_{i(j)}$ represent the masses of $i \, (j)$ channel of the incoming (outgoing) particles, and the coefficients of $C_{ij}$ can be found in Ref. \cite{Geng:2006yb} (also given in Appendix \ref{appb}). Then, we can input these potentials into the the on-shell Bethe-Salpeter equation to evaluate the scattering amplitude,
\begin{equation}
t^I  =  [1+V^I \hat{G}^I ]^{-1} (-V^I) \vec{\epsilon} \cdot \vec{\epsilon}\;',
\end{equation}
where $\hat{G}^I$ is $(1+ \frac{1}{3}\frac{q_l^2}{M_l^2})G^I$ being a diagonal matrix ($G^I$ as the normal loop function in Eq. \eqref{eq:bse}) and $\vec{\epsilon} (\vec{\epsilon}\;') $ represents a polarization vector of the incoming (outgoing) vector-meson. As done in Ref. \cite{Geng:2006yb} (details of the convolution of $G^I$ given in Appendix \ref{appc}), we also take into account the large width of the vector mesons, and consider the convolution of the vector mesons as intermediate state in the loop function $G^I$. In Fig. \ref{fig:trhoK}, we show our results for the modulus squared of $t_{\rho K}^{I=1/2}$, which are consistent with Ref. \cite{Geng:2006yb}. From the peak position, we can see that the $K_1(1270)$ is successfully reproduced in our work as the cluster in the FCA. In $I=3/2$ sector, there are only two coupled channels, $\rho K$ and $K^* \pi$, which is no resonance appeared. But for considering the isospin structure of the subsystem, we also evaluate the amplitude $t_{\rho K}^{I=3/2}$, which is not showed in the figure.

\section{Results}
\label{secres}

In the present work, we study the $K$-multi-$\rho$ interactions. In the former section, we have reproduced the resonances $f_2(1270)$ and $K_1(1270)$ in the $\rho \rho$ and $\rho K$ two-body interactions, which are the clusters of the FCA to Faddeev equations for the three-body interaction. Therefore, based on the possible clusters in the two-body interaction discussed above, the possible cases for the $K$-multi-$\rho$ interactions are listed in Table \ref{tab:cases}, and explained as follows. For the three-body interaction, we have two options: (i) particle $3=K$, cluster or resonance $R=f_2$ (particle $1=\rho,\;2=\rho$) and (ii) $3=\rho$, $R=K_1$ ($1=\rho,\;2=K$). For the four-body interaction, we can extrapolate the FCA ideas and also have two cases: (i) $3=f_2$, $R=K_1$ ($1=\rho,\;2=K$) and (ii) $3=K_1$, $R=f_2$ ($1=\rho,\;2=\rho$). If we find a new resonance in the four-body interaction, assumed as $K_3$, thus, there are also two cases for the five-body interaction, (i) $3=K$, $R=f_4$ ($1=f_2,\;2=f_2$) and (ii) $3=\rho$, $R=K_3$ ($1=f_2,\;2=K_1$). For the six-body interaction, (i) $3=K_1$, $R=f_4$ ($1=f_2,\;2=f_2$) and (ii) $3=f_2$, $R=K_3$ ($1=f_2,\;2=K_1$). We show our investigation results for all these cases as below.
\begin{table}[htb]
\centering
\caption{The cases considered in the $K$-multi-$\rho$ interactions. Threshold 1: $m_3 + m_1 + m_2$; threshold 2: $m_3 + M_R$ (unit: MeV).}
\label{tab:cases}
\begin{tabular}{cccccc}
\hline\hline
\hspace{0.3cm} particles: \hspace{0.3cm} & \hspace{0.5cm} 3 \hspace{0.5cm} & \hspace{0.5cm} R (1,2) \hspace{0.5cm} &  \hspace{0.3cm} amplitudes \hspace{0.3cm} & \hspace{0.3cm} threshold 1 \hspace{0.3cm} & \hspace{0.3cm} threshold 2 \hspace{0.3cm} \\
\hline
  & $\rho$ & $K$ &  $t_{\rho K}$ & 1271.0 & (1270.0)$_{K_1}$ \\
\rb{Two-body}  & $\rho$ & $\rho$ & $t_{\rho\rho}$ & 1551.0 & (1275.1)$_{f_2}$ \\
\hline
  & $K$ & $f_2\;(\rho\rho)$ & $T_{K-f_2}$ & & 1770.6 \\
\rb{Three-body}  & $\rho$ & $K_1\;(\rho K)$ & $T_{\rho-K_1}$ & \rb{2046.5} & 2045.5 \\
\hline
  & $K_1$ & $f_2\;(\rho\rho)$ & $T_{K_1-f_2}$ & 2821.0 &  \\
\rb{Four-body}  & $f_2$ & $K_1\;(\rho K)$ & $T_{f_2-K_1}$ & 2546.1 & \rb{2545.1}\\
\hline
  & $K$ & $f_4\;(f_2 f_2)$ & $T_{K-f_4}$ & 3045.7 & 2513.5  \\
\rb{Five-body}  & $\rho$ & $K_3\;(f_2 K_1)$ & $T_{\rho-K_3}$ & 3320.6 & (2855.5)  \\
\hline
  & $K_1$ & $f_4\;(f_2 f_2)$ & $T_{K_1-f_4}$ & & 3288 \\
\rb{Six-body}  & $f_2$ & $K_3\;(f_2 K_1)$ & $T_{f_2-K_3}$ & \rb{3820.2} & (3355.5) \\
\hline\hline
\end{tabular}
\end{table}

\subsection{Three-body interaction}
\label{threeb}

In the three-body interaction, there are two possible structures: $K-f_2(\rho\rho)$ and $\rho-K_1(\rho K)$, which means (i) $3=K$, $R=f_2$ ($1=\rho,\;2=\rho$) and (ii) $3=\rho$, $R=K_1$ ($1=\rho,\;2=K$). To evaluate these scattering amplitudes, we need as input the $t_1$ and $t_2$ amplitudes of the (3,1) and (3,2) subsystems, $t_1 = t_2 = t_{\rho K}$ for $K-f_2(\rho\rho)$ and $t_1 = t_{\rho\rho},\; t_2=t_{\rho K}$ for $\rho-K_1(\rho K)$. We can calculate them in the two-body $\rho\rho$ and $\rho K$ interactions, which are discussed in the former section, following the work of Refs. \cite{Molina:2008jw,Geng:2006yb}. But, note that, in their work, the dimensional regularization scheme is used for the loop functions. To evaluate the form factor of the cluster, we need a cutoff $\Lambda'$, which is the same as the one $q_{max}$ used in the loop function for the two-body interaction, discussed in Section \ref{secform}. As discussed in Ref. \cite{Xiao:2011rc}, we can compare the value of the $G$ function at threshold using the dimensional regularization formula \cite{Oset:2001cn} with the one of the cut off which can be taken from Ref. \cite{Oller:1997ti} or the analytic expression in Ref. \cite{Guo:2005wp}. Then, equivalently to the parameters in the dimensional expression, we obtain $q_{max}=875\mev$ for the $f_2(1270)$ cluster and $q_{max}=1035\mev$ for the $K_1(1270)$ cluster. In fact, we do not introduce any free parameter.

\begin{figure}
\centering
\includegraphics[scale=0.6]{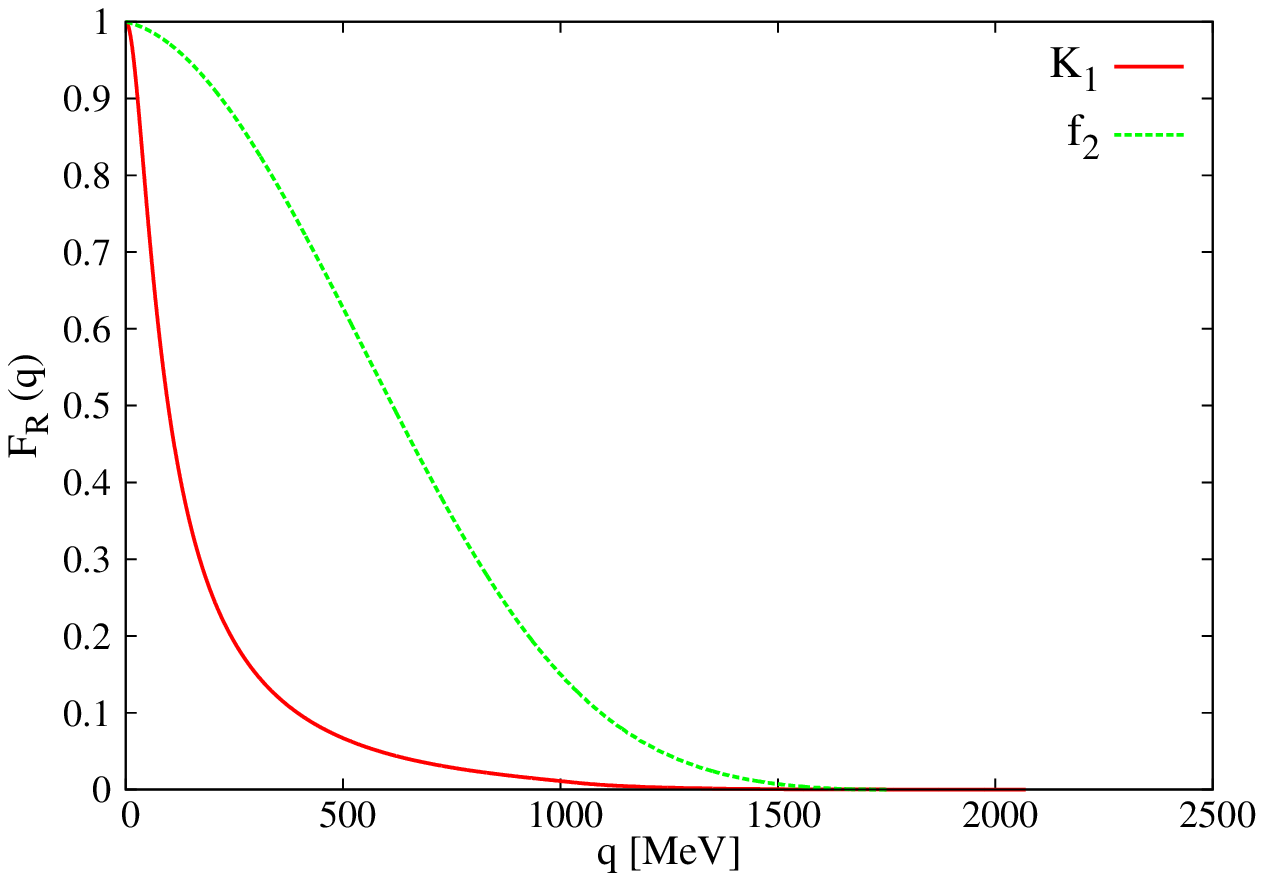}
\includegraphics[scale=0.6]{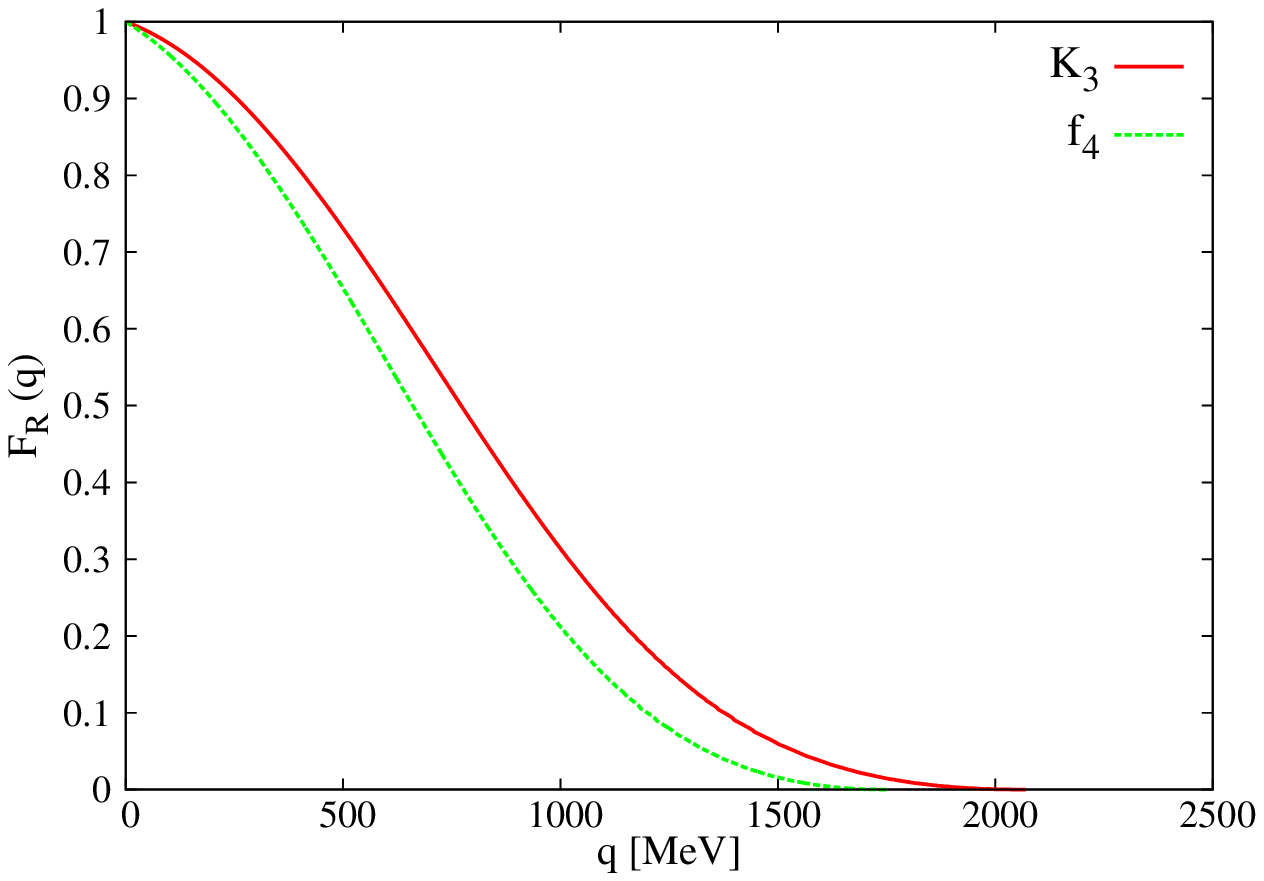}
\caption{The form factors of $f_2(1270)$ and $K_1(1270)$ (Left), $f_4(2050)$ and $K_3(2090)$ (right).}
\label{fig:formf_K1f2}
\end{figure}

With the values of these two $q_{max}$ for the cutoff of $\Lambda'$, then using Eqs. \eqref{eq:formfactor} and \eqref{eq:formfactorN}, we can evaluate the form factors of the clusters, $f_2(1270)$ and $K_1(1270)$, shown in Fig. \ref{fig:formf_K1f2} (Left), where we can see that when $q \to 2 \Lambda', \ F_R (q) \to 0$. Therefore, for the cutoff of the $G_0$ function, seen in Eq. \eqref{eq:G0s}, we choose as $2 \Lambda'$ from the constraint of the form factor. Note that, the form factor of Eq. \eqref{eq:formfactor} is only valid for the cluster of bound state, which is the starting point of the FCA, since only the wave functions of the components of the bound state lead to the ordinary form factor but not for the cluster of resonance \cite{YamagataSekihara:2010pj} (this is the limitation of the FCA, as discussed in the formalism). In fact, we also have two more form factors for the clusters, the $f_4$ and $K_3$ states, seen in Table \ref{tab:cases}. Since the form factor often reduces fast above a certain momentum ($q \simeq 2 \Lambda'$), as shown in Fig. \ref{fig:formf_K1f2} (Left), we can safely choose the same cutoffs as the $f_2$ and $K_1$ for the ones of the $f_4$ and $K_3$ respectively to avoid introducing any new free parameters, seen in Fig. \ref{fig:formf_K1f2} (Right). Furthermore, even though we change the value of the determined $q_{max}$ a bit, for example 10\%, if it is still in the nature value \cite{Oller:2000fj}, the shape of the form factor will not be changed, seen Fig. \ref{fig:formf_K1f2}, thus, our final conclusion will not be changed, and only the strength of the amplitude will be a little differences.

We have mentioned in the formalism that we should take into account the isospin structure of the subsystems for the two-body amplitudes $t_1$ and $t_2$. For the first case of $K-f_2(\rho\rho)$, the cluster of $f_2$ resonance has isospin $I=0$. Therefore the two $\rho$ mesons are in an $I=0$ state, and we write it in terms of the physical basis components
\begin{equation} 
|\rho\rho>^{(0,0)}=\frac{1}{\sqrt{3}} \Big( |(1,-1)> \;+\; |(-1,1)> \;-\; |(0,0)> \Big),
\end{equation}
where $|(1,-1)>$ denote $|(I_z^1,I_z^2)>$ which shows the $I_z$ components of particles 1 and 2, and $|\rho\rho>^{(0,0)}$ means $|\rho\rho>^{(I,I_z)}$. Then, the third particle is a $K$ meson taken $|I_z^3>=|\frac{1}{2}>$, obtained
\begin{equation} 
\begin{split}
T^{(\frac{1}{2},\frac{1}{2})}_{K-f_2}=&<K - \rho\rho|\,\hat{t}\,|K - \rho\rho>^{(\frac{1}{2},\frac{1}{2})}  \\
=&(<K|^{(\frac{1}{2},\frac{1}{2})} \otimes <\rho\rho|^{(0,0)})\,(\hat{t}_{31}+\hat{t}_{32})\,(|K>^{(\frac{1}{2},\frac{1}{2})} \otimes |\rho\rho>^{(0,0)})\\
=&\Big[ <\frac{1}{2}| \otimes \frac{1}{\sqrt{3}} \Big( <(1,-1)| + <(-1,1)| - <(0,0)| \Big) \Big]\,(\hat{t}_{31}+\hat{t}_{32})\,\Big[ |\frac{1}{2}> \\
&\otimes \frac{1}{\sqrt{3}} \Big( |(1,-1)> + |(-1,1)> - |(0,0)> \Big) \Big]\\
=&\frac{1}{3} \Big[ <(\frac{3}{2},\frac{3}{2}),-1| + \sqrt{\frac{1}{3}} <(\frac{3}{2},-\frac{1}{2}),1| + \sqrt{\frac{2}{3}} <(\frac{1}{2},-\frac{1}{2}),1| - \sqrt{\frac{2}{3}} <(\frac{3}{2},\frac{1}{2}),0| \\
&- \sqrt{\frac{1}{3}} <(\frac{1}{2},\frac{1}{2}),0| \Big] \hat{t}_{31} \Big[ |(\frac{3}{2},\frac{3}{2}),-1> + \sqrt{\frac{1}{3}} |(\frac{3}{2},-\frac{1}{2}),1> + \sqrt{\frac{2}{3}} |(\frac{1}{2},-\frac{1}{2}),1> \\
&- \sqrt{\frac{2}{3}} |(\frac{3}{2},\frac{1}{2}),0> - \sqrt{\frac{1}{3}} |(\frac{1}{2},\frac{1}{2}),0> \Big] +\frac{1}{3} \Big[ \sqrt{\frac{1}{3}} <(\frac{3}{2},-\frac{1}{2}),1| + \sqrt{\frac{2}{3}} <(\frac{1}{2},-\frac{1}{2}),1| \\
&+ <(\frac{3}{2},\frac{3}{2}),-1| - \sqrt{\frac{2}{3}} <(\frac{3}{2},\frac{1}{2}),0| - \sqrt{\frac{1}{3}} <(\frac{1}{2},\frac{1}{2}),0| \Big] \hat{t}_{32} \Big[  \sqrt{\frac{1}{3}} |(\frac{3}{2},-\frac{1}{2}),1> \\
&+ \sqrt{\frac{2}{3}} |(\frac{1}{2},-\frac{1}{2}),1> + |(\frac{3}{2},\frac{3}{2}),-1> - \sqrt{\frac{2}{3}} |(\frac{3}{2},\frac{1}{2}),0> - \sqrt{\frac{1}{3}} |(\frac{1}{2},\frac{1}{2}),0> \Big],
\label{eq:tKf2}
\end{split}
\end{equation}
where the notation of the states followed in the terms is $|(\frac{3}{2},\frac{3}{2}),-1> \equiv |(I^{31},I_z^{31}),I_z^2>$ for $t_{31}$, and $|(I^{32},I_z^{32}),I_z^1>$ for $t_{32}$. Finally, we obtain the amplitudes combining with the isospin structure
\begin{equation}
t_1 = t_{\rho K} = \frac{1}{3} \big( 2 \;t_{31}^{I=3/2} + t_{31}^{I=1/2} \big),\quad t_2 = t_1, \label{eq:tKrho}
\end{equation}
where $\rho K$ scattering amplitudes with isospin $I = 1/2, \, 3/2$, $t_{31}^{I=1/2}$ and $t_{31}^{I=3/2}$, have been evaluated in Subsection \ref{subsecRK}.

But for the second case of $\rho-K_1(\rho K)$, the isospin structure relationship is different. Now, the isospins of $\rho$ and $K_1$ are $I_\rho=1$ and $I_{K_1}=\frac{1}{2}$, thus, the total isospin of the three-body system are two cases $I_{total}\equiv I_{\rho\rho K}=\frac{1}{2}$ or $I_{total}\equiv I_{\rho\rho K}=\frac{3}{2}$. Therefore we have
\begin{equation} 
\begin{split}
&|\rho K_1>^{(\frac{1}{2},\frac{1}{2})} = |\rho\rho K>^{(\frac{1}{2},\frac{1}{2})} = \sqrt{\frac{2}{3}} \;|(1,-\frac{1}{2})> \;-\; \sqrt{\frac{1}{3}} \;|(0,\frac{1}{2})>, \\
&|\rho K_1>^{(\frac{3}{2},\frac{1}{2})} = |\rho\rho K>^{(\frac{3}{2},\frac{1}{2})} = \sqrt{\frac{1}{3}} \;|(1,-\frac{1}{2})> \;+\; \sqrt{\frac{2}{3}} \;|(0,\frac{1}{2})>,
\label{eq:trhoK1}
\end{split}
\end{equation}
where we have taken the third isospin component $I_z = \frac{1}{2}$ for convenience. Then the $|\rho K>$ states inside the $K_1$ for the $I_z = -\frac{1}{2}$ and $I_z = +\frac{1}{2}$ are given by
\begin{equation} 
\begin{split}
&|\rho K>^{(\frac{1}{2},-\frac{1}{2})} = \sqrt{\frac{1}{3}} \;|(0,-\frac{1}{2})> \;-\; \sqrt{\frac{2}{3}} \;|(-1,\frac{1}{2})>, \\
&|\rho K>^{(\frac{1}{2},\frac{1}{2})} = \sqrt{\frac{2}{3}} \;|(1,-\frac{1}{2})> \;-\; \sqrt{\frac{1}{3}} \;|(0,\frac{1}{2})>.
\label{eq:trhoK2}
\end{split}
\end{equation}
For the two possibilities, using Eqs. \eqref{eq:trhoK1} and \eqref{eq:trhoK2} and performing a similar derivation of Eq. \eqref{eq:tKf2}, we obtain
\begin{equation}
\begin{split}
&T_{\rho-K_1}^{(I=1/2)}: \quad t_1 = t_{\rho\rho} = \frac{2}{3} \;t_{31}^{I=0}, \quad t_2 = t_{\rho K} = \frac{1}{9} \big( 8 \;t_{32}^{I=3/2} + t_{32}^{I=1/2} \big); \\
&T_{\rho-K_1}^{(I=3/2)}: \quad t_1 = t_{\rho\rho} = \frac{5}{6} \;t_{31}^{I=2}, \quad t_2 = t_{\rho K} = \frac{1}{9} \big( 5 \;t_{32}^{I=3/2} + 4 \;t_{32}^{I=1/2} \big),
\end{split}
\end{equation}
where $\rho \rho$ interaction amplitudes with isospin $I = 0, \, 2$, $t_{31}^{I=0}$ and $t_{31}^{I=2}$, are evaluated in Subsection \ref{subsecRR}, and the ones of $\rho K$ interaction, $t_{32}^{I=1/2}$ and $t_{32}^{I=3/2}$, are given in Subsection \ref{subsecRK}.

\begin{figure}
\centering
\includegraphics[scale=0.6]{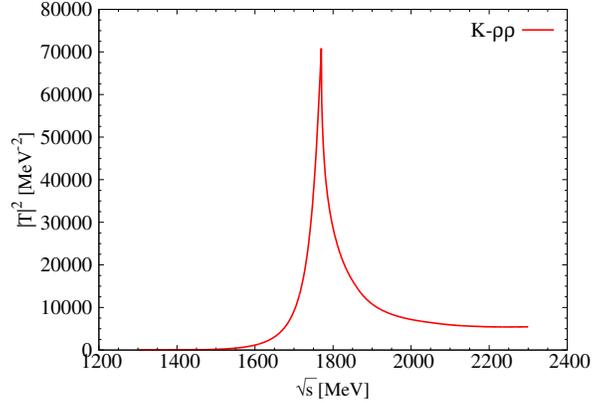}
\caption{Modulus squared of the $T_{K-f2}$.}
\label{fig:t3krr}
\end{figure}

\begin{figure}
\centering
\includegraphics[scale=0.6]{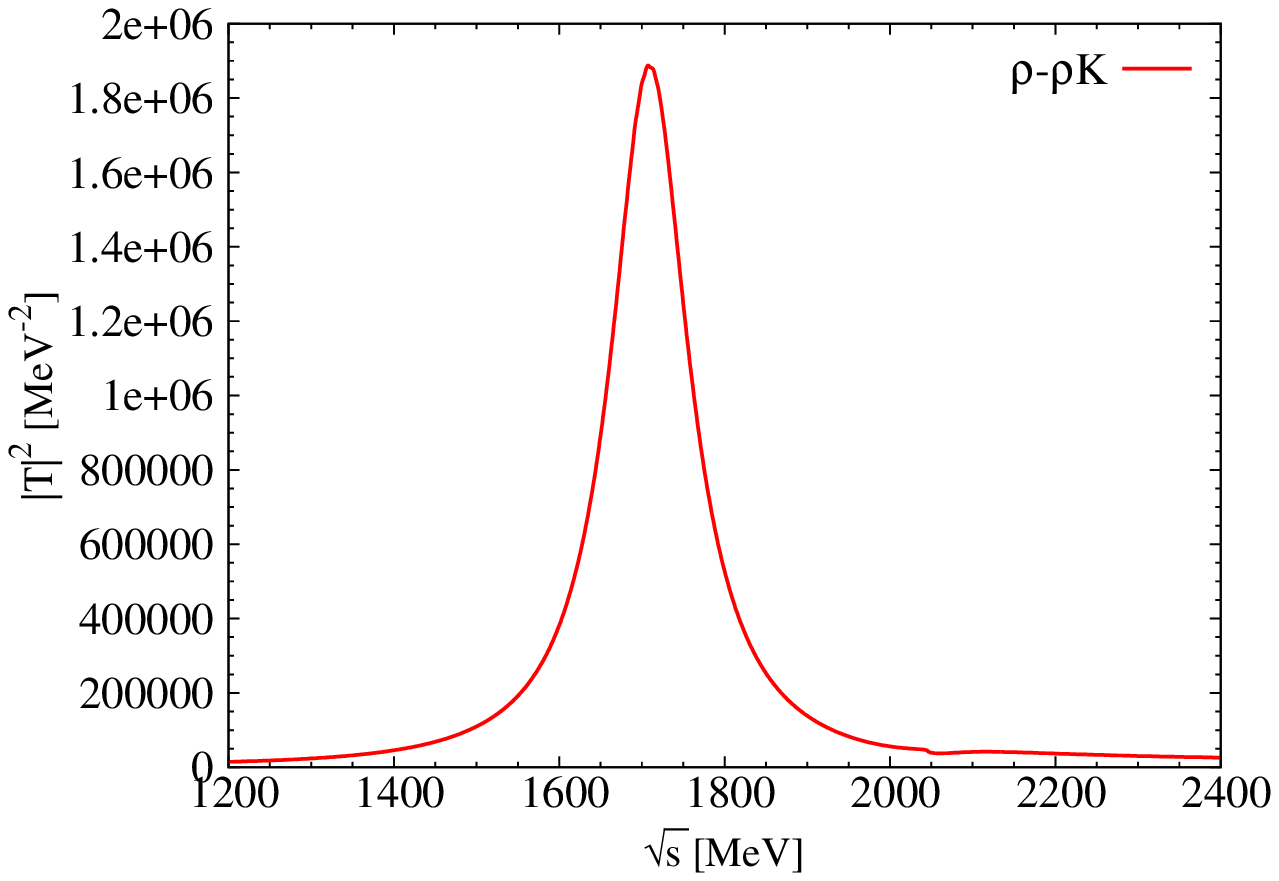}
\includegraphics[scale=0.6]{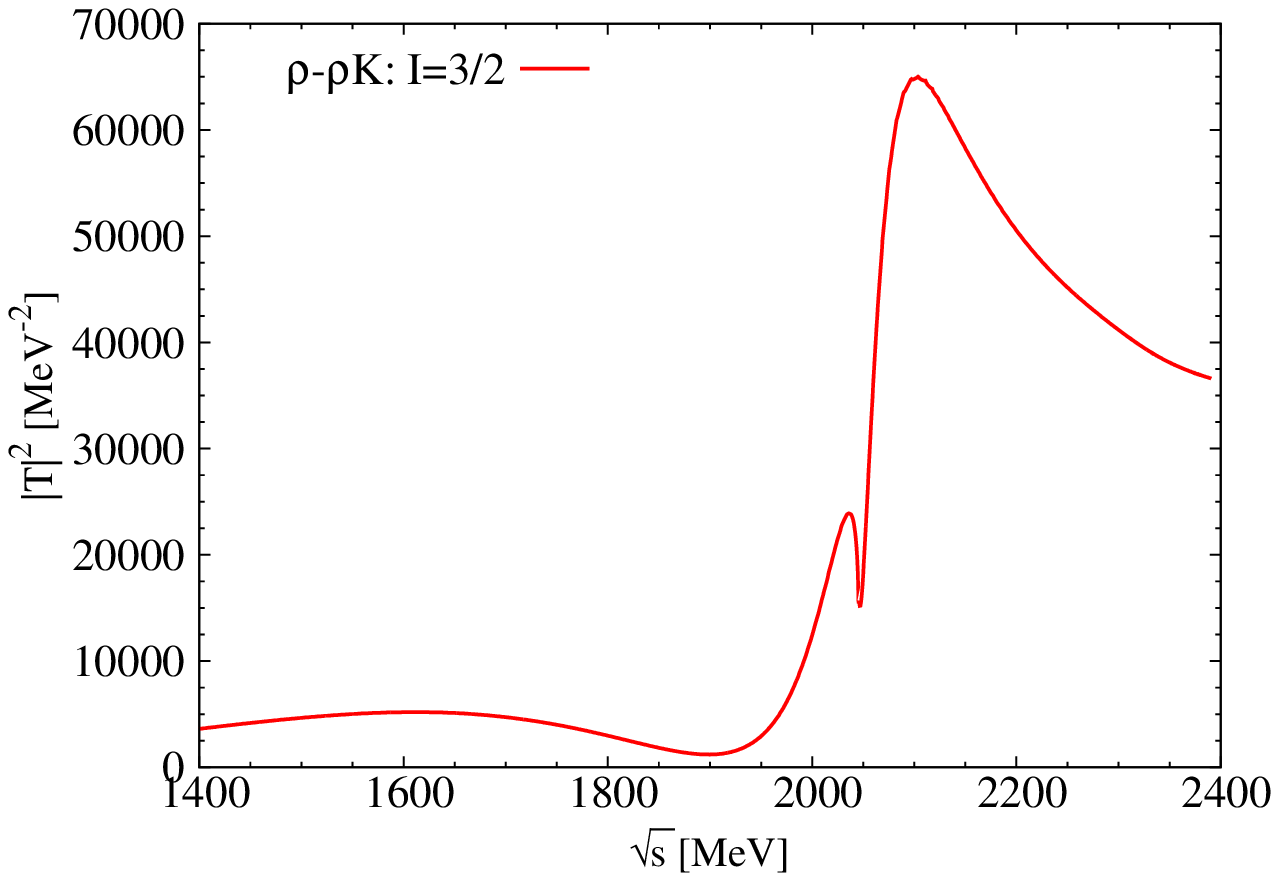}
\caption{Modulus squared of the $T_{\rho-K_1}$ scattering amplitudes. Left: $I_{total}=\frac{1}{2}$; Right: $I_{total}=\frac{3}{2}$.} \label{fig:t3rrk}
\end{figure}

In Fig. \ref{fig:t3krr} we show our results of the modulus squared of the amplitude for $|T_{K-f_2}^{I=1/2}|^2$. There is a clear and sharp peak around $1770\mev$, which is close to the threshold of $K-f_2$ and similar to a cusp in the case of the $\eta' K \bar{K}$ system \cite{Liang:2013yta}. Therefore, this peak in $|T_{K-f_2}^{I=1/2}|^2$ would be affected by the threshold effect, and the width of this resonance structure has large uncertainties. There would be the cusp corresponded to a real resonance in some cases, like $a_0 (980)$ \cite{Oller:1997ti} (more discussions about the states appearing near the threshold can be found in the recent works of \cite{Guo:2014iya,Hyodo:2014bda}). In Fig. \ref{fig:t3rrk} we show the results of $|T_{\rho-K_1}^{I=1/2}|^2$ (left) and $|T_{\rho-K_1}^{I=3/2}|^2$ (right). From the $|T_{\rho-K_1}^{I=1/2}|^2$ results, we find that there is a clear peak around the energy $1707\mev$ with a width about $113\mev$ \footnote{Since the complicate situation in the definition of the second Riemann sheets in the multi-body interaction, we have the difficulty to extract the masses by the poles in the second Riemann sheets, which is different from the two-body interaction, more discussions seen in Ref.~\cite{Bayar:2013bta}. Thus, we extract the mass and the width from the scattering amplitude in the first Riemann sheet, shown in the figures.}, which is about $340\mev$ below the $\rho-K_1$ threshold and the $\rho \rho K$ threshold. Because of the large width of the $\rho$ meson, for a system with two $\rho$ mesons, the large bindings in the present case will be acceptable. In PDG \cite{pdg2014}, the $K_2 (1770)$ of $J^P=2^-$ strangeness state is the mass of $1773 \pm 8 \mev$ and the width $186 \pm 14 \mev$. But, from the analysis of the $K \omega$ spectrum in the reaction $K^- p \to K^- \omega p$, the work of Ref. \cite{Chung:1974wb} obtains its mass as $1710 \pm 15 \mev$ and width $110 \pm 50 \mev$, which is consistent with our results. Our results are also consistent with the other experimental results \cite{AguilarBenitez:1970up,Blieden:1972aj,Tikhomirov:2003gg}. Thus, considering the uncertainties in our study (which will be discussed later), the peak appearing in the $|T_{\rho-K_1}^{I=1/2}|^2$ corresponds to the $K_2 (1770)$, which would be the $\rho-K_1$ molecular state in our model. The strength of the peak of $|T_{K-f_2}^{I=1/2}|^2$ is about 25 times smaller than for $|T_{\rho-K_1}^{I=1/2}|^2$, thus, we could not expect a state structure in Fig. \ref{fig:t3krr}, even though the two-body interaction of $\rho K$ is not strong as $\rho \rho$ by comparing the results of Fig. \ref{fig:trhorho} and Fig. \ref{fig:trhoK}. From the results of $|T_{\rho-K_1}^{I=3/2}|^2$ in Fig. \ref{fig:t3rrk} (right), we can see that there is a clear resonant structure about $2100\mev$ with the strength 25 times smaller than that of $|T_{\rho-K_1}^{I=1/2}|^2$ in the left figure, which is a little above the $\rho-K1(1270)$ threshold, and also show a dip in the threshold. We are looking for the lowest lying states bound in the $K$-multi-$\rho$ system, and hence, we could not expect a new state in the $|T_{\rho-K_1}^{I=3/2}|^2$ results.

\subsection{Four-body interaction}
\label{fourb}

For the four-body interaction as shown in Table \ref{tab:cases}, we also have two possibilities: (i) particle $3=f_2$, cluster $R=K_1$ ($1=\rho,\;2=K$), or (ii) particle $3=K_1$, resonance $R=f_2$ ($1=\rho,\;2=\rho$). Because the isospins of the two cluster are $I_{f_2} = 0$ and $I_{K_1} = \frac{1}{2}$, the total isospin of the four-body system is only $I_{total} = \frac{1}{2}$. Using the FCA formalism as discussed in Section \ref{secform} for the four-body interaction, for the first option, $f_2$ interacting with the $K_1$, we need to evaluate the amplitudes $t_1 = t_{f_2 \rho} = T_{\rho-f_2}$, which has been done in Ref. \cite{Roca:2010tf} \footnote{where they used a similar formalism of Sec. \ref{secform}, thus, using the formalism of Sec. \ref{secform} and the $\rho \rho$ interaction amplitudes in Subsec. \ref{subsecRR}, we can reproduce their results.}, and $t_2 = t_{f_2 K} = T_{K-f_2}$, which has been calculated in the former Subsection \ref{threeb}. Similarly, for the second case, $K_1$ collides with the $f_2$, and the amplitudes $t_1 = t_2 = t_{K_1 \rho} = T_{\rho-K_1}$ have been evaluated in the former Subsection \ref{threeb}. Note that the three-body amplitude $T_{\rho-K_1}$ should be also written in terms of the isospin structure as discussed in the Section \ref{secform} and its amplitudes of isospin components evaluated in the last Subsection \ref{threeb}. Since the isospins of both the $K_1$ and $K$ are $I=\frac{1}{2}$, the isospin structure is similar to the case when the $K$ collides with the $f_2$. Thus, analogously to Eq. \eqref{eq:tKrho} we have
\begin{equation}
t_1 = T_{\rho K_1} = \frac{1}{3} \big( 2 T_{31}^{I=3/2} + T_{31}^{I=1/2} \big),\quad t_2 = t_1.\label{eq:threet3}
\end{equation}

\begin{figure}
\centering
\includegraphics[scale=0.6]{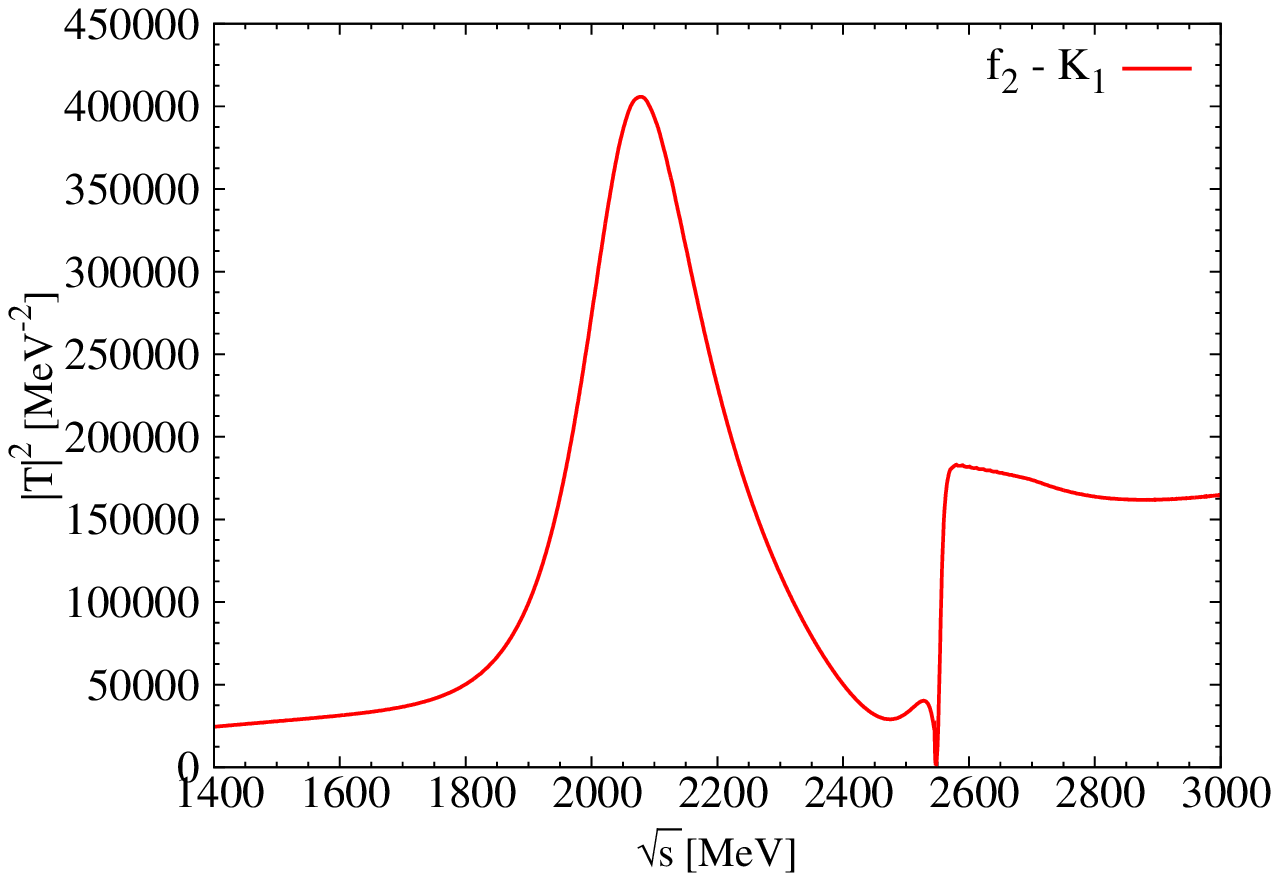}
\includegraphics[scale=0.6]{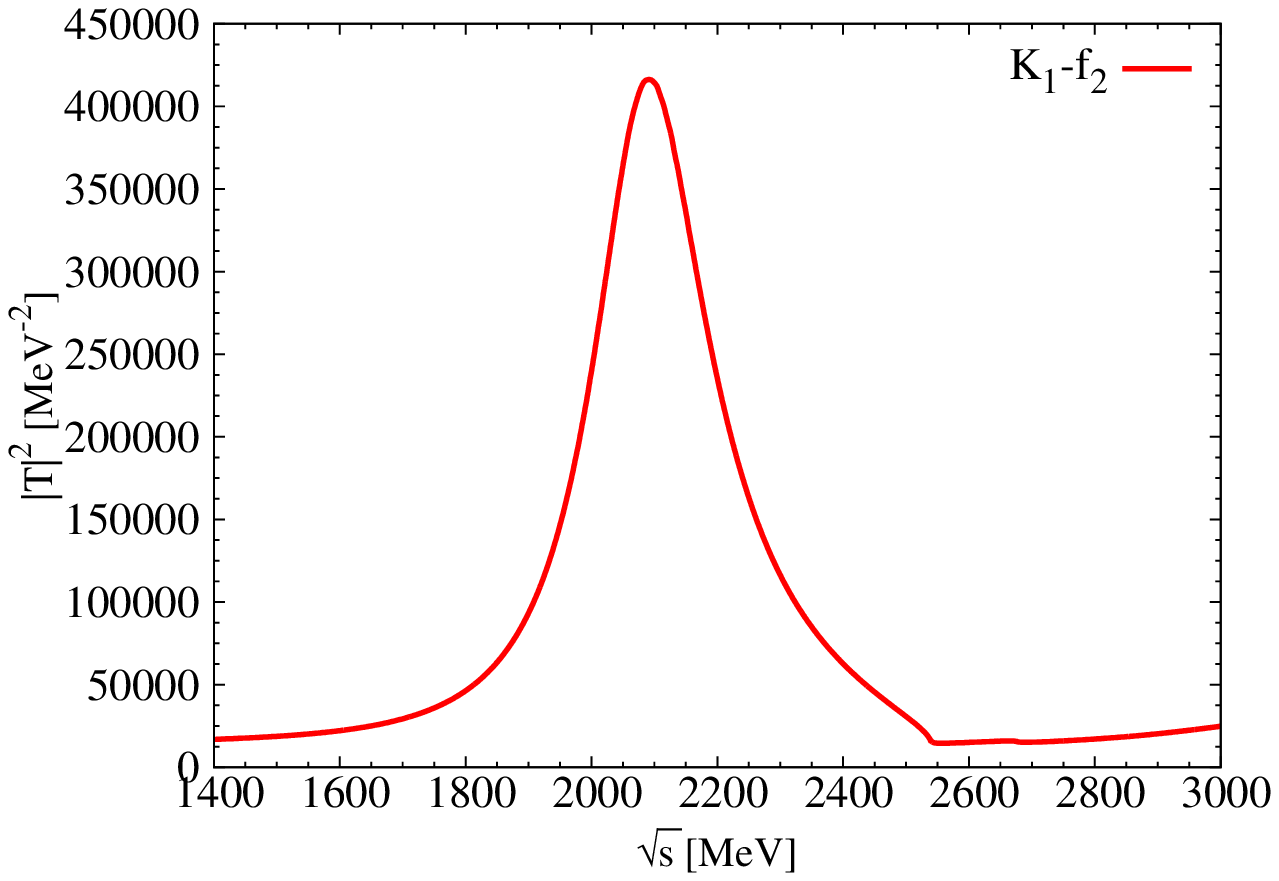}
\caption{Modulus squared of the $T_{f_2-K_1}$ (left) and $T_{K_1-f2}$ (right) scattering amplitudes.}
\label{fig:t4k1f2}
\end{figure}

We show our results in Fig. \ref{fig:t4k1f2}, of which the left is $|T_{f_2-K_1}^{I=1/2}|^2$ and the right $|T_{K_1-f_2}^{I=1/2}|^2$, where the clusters are resonances $K_1$  and $f_2$ respectively. In the left of Fig. \ref{fig:t4k1f2}, there is a clear peak at an energy of about $2079\mev$, the width of which is about $249\mev$. We also find discontinuity at the position about $2550\mev$, which comes from the contribution of the $f_2-K_1$ threshold effect. In the right of Fig. \ref{fig:t4k1f2}, we also find that there is a resonant peak around the energy $2091\mev$ with a width of $230\mev$ in the $|T_{K_1-f_2}^{I=1/2}|^2$. The strength of the peak of $|T_{K_1-f_2}^{I=1/2}|^2$ is at the same magnitude as the one of $|T_{f_2-K_1}^{I=1/2}|^2$, and the energy of the peak is a little less bound. In PDG, there is only one $J^P = 3^+ \ K_3$ state found in the experiments, $K_3 (2320)$, with a mass $2324 \pm 24 \mev$ and width $150 \pm 30 \mev$. This $K_3 (2320)$ state was just found in the reactions $K^+ p \to (\bar{\Lambda} p)p$ and $K^- p \to (\Lambda \bar{p})p$ \cite{Cleland:1980ya,Armstrong:1983gt}. But, its mass is too far away from our results. Therefore, we find a new $K_3$ state with uncertainties, with a mass about $2079 \sim 2091\mev$ and a width about $230 \sim 249\mev$.

\subsection{Five-body interaction}

As we expected before, we also find a new $K_3$ state in the four-body interaction in the former Subsection \ref{fourb}. Thus, for the five-body interaction, there also are two options for the cluster, one of which is the $f_4$ state found in the PDG and studied in Ref. \cite{Roca:2010tf}, and the other one the resonance $K_3$ obtained in the four-body interaction above. Then following the idea of FCA and letting the third particle ($K$ or $\rho$) collide with them, the two possibilities are (i) particle $3=K$, cluster $R=f_4$ ($1=f_2,\;2=f_2$), or (ii) $3=\rho$, $R=K_3$ ($1=f_2,\;2=K_1$). Since the isospin $I_{f_4} = 0$ and $I_{K_3} = \frac{1}{2}$, for the first case, the total isospin of the $K-f_4$ system is only $I_{total} = \frac{1}{2}$, but for the second option $\rho-K_3$, the total isospin of this structure is $I_{total} = \frac{1}{2}$ or $I_{total} = \frac{3}{2}$. Therefore, the situation of $K$ interacting with $f_4$ ($K-f_4$) is similar to the three-body interaction discussed before, $K$  colliding with $f_2$ ($K-f_2$), and $\rho-K_3$ analogous to the one of $\rho-K_1$. Thus, in the first case, the $K$ collides with the $f_4$, and the amplitudes $t_1 = t_2 = t_{K f_2} = T_{K-f_2}^{(I=1/2)}$ have been evaluated in Subsection \ref{threeb} for the three-body interaction. For the second case, the $\rho$ interacts with the $K_3$, which is similar to $\rho-K_1$ in Subsection \ref{threeb}, thus, doing a similar derivation as done in Eq. \eqref{eq:tKf2}, we obtain
\begin{equation} 
\begin{split}
&T_{\rho-K_3}^{(I=1/2)}: \quad t_1 = t_{\rho f_2} = T_{31}^{(I=1)}, \quad t_2 = t_{\rho K_1} =  T_{32}^{I=1/2}; \\ 
&T_{\rho-K_3}^{(I=3/2)}: \quad t_1 = t_{\rho f_2} = T_{31}^{(I=1)}, \quad t_2 = t_{\rho K_1} =  T_{32}^{I=3/2},
\end{split}
\end{equation}
where the $T_{31}^{(I=1)}$ is the amplitude of $T_{\rho-f_2}$, which is the same as calculated in the Subsection \ref{fourb} reproducing the results of Ref. \cite{Roca:2010tf}, and the amplitudes $T_{\rho-K_1}^{I=1/2}$ and $T_{\rho-K_1}^{I=3/2}$ have also been evaluated in Subsection \ref{threeb}.

\begin{figure}
\centering
\includegraphics[scale=0.6]{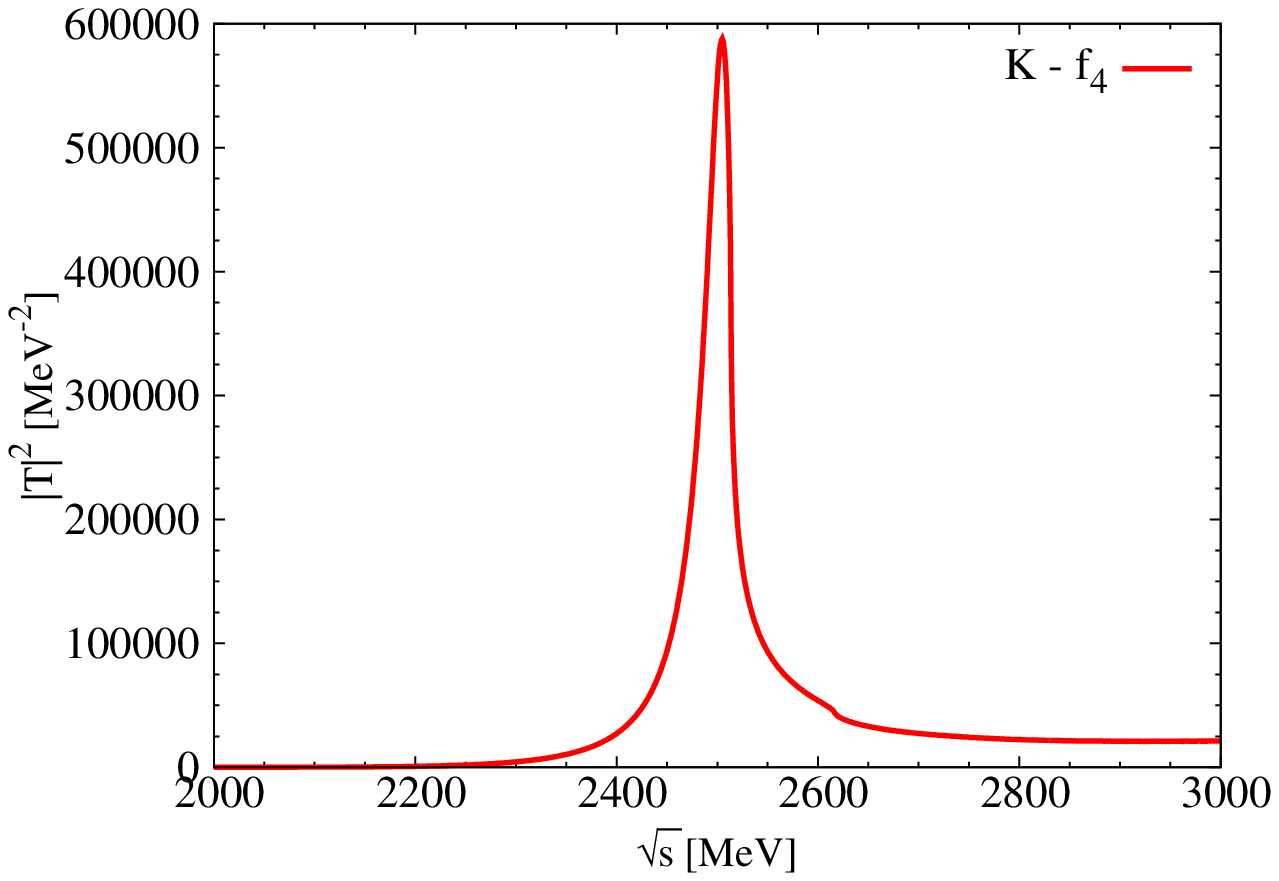}
\includegraphics[scale=0.6]{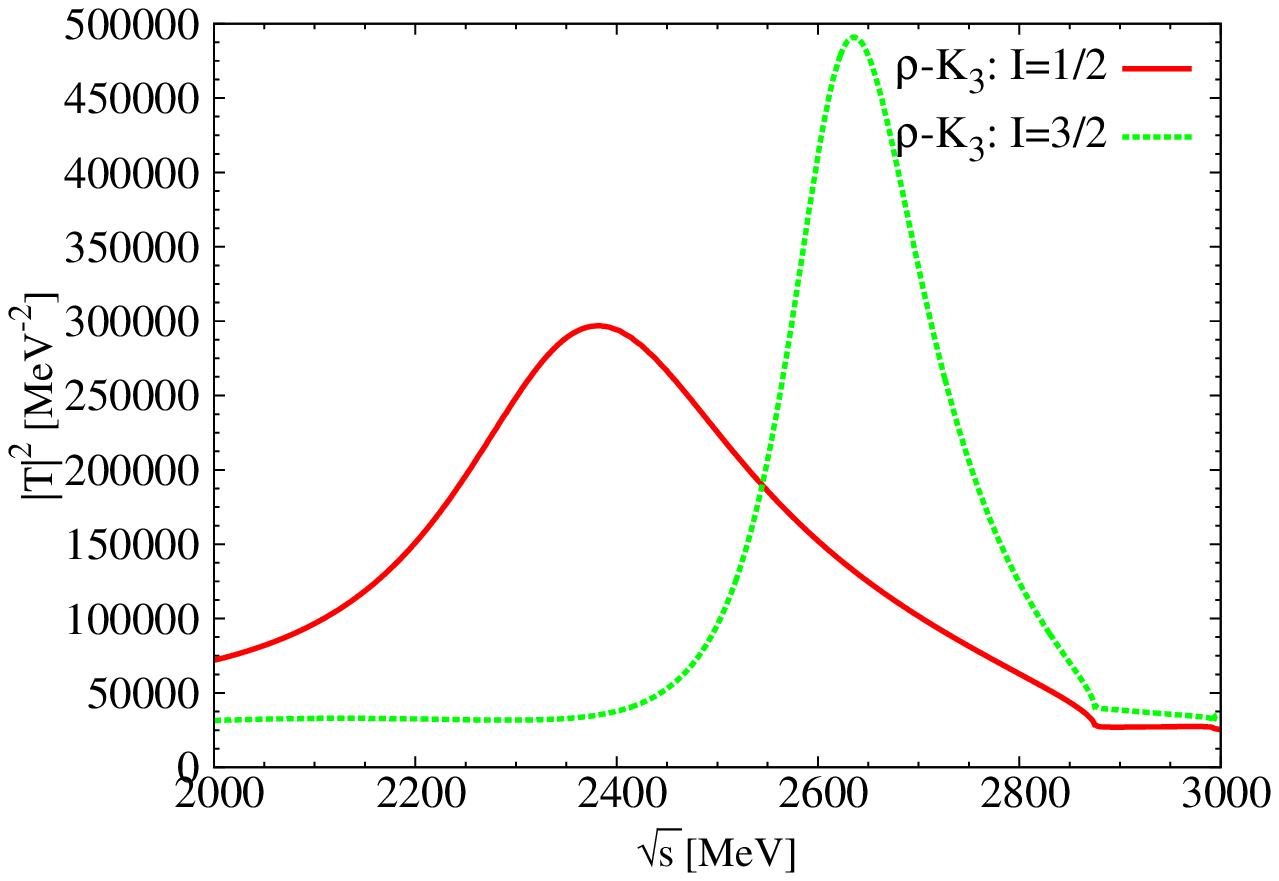}
\caption{Modulus squared of the $T_{K-f_4}$ (left) and $T_{\rho-K_3}$ (right) scattering amplitudes.}
\label{fig:t5kf4rk3}
\end{figure}

We show our results for the two cases of the five-body interaction in Fig. \ref{fig:t5kf4rk3}. The results of $|T_{K-f_4}^{I=1/2}|^2$ is shown on the left of Fig. \ref{fig:t5kf4rk3}, where we can see a resonant peak around the energy $2505\mev$ with a width of about $32\mev$. The right of Fig. \ref{fig:t5kf4rk3} is the results of $|T_{\rho-K_3}^{I=1/2}|^2$ and $|T_{\rho-K_3}^{I=3/2}|^2$. We observe that there are clear peaks for both of them. A resonant structure in $|T_{\rho-K_3}^{I=1/2}|^2$ is found at the energy $2382\mev$ with the width about $409\mev$, which is about $120\mev$ more bound than the one of $|T_{K-f_4}^{I=1/2}|^2$ and much larger width too. But the strength of $|T_{\rho-K_3}^{I=1/2}|^2$ is just one half smaller than the one $|T_{K-f_4}^{I=1/2}|^2$. Thus, the more bound energy and lager width in $|T_{\rho-K_3}^{I=1/2}|^2$ come from the stronger $\rho \rho$ interaction in the components of $K_3$ state, where we can see from the Fig. \ref{fig:trhorho} and Fig. \ref{fig:trhoK}. In experiments, one $J^P = 4^-$ state was also found in Ref. \cite{Cleland:1980ya}, $K_4 (2500)$, with the mass $2490 \pm 20 \mev$ and width around $250 \mev$, which is not well confirmed in PDG because of the lack of more experimental information. This $K_4 (2500)$ state is dynamically generated as a molecular state of $K-f_4$ in our results of $|T_{K-f_4}^{I=1/2}|^2$, with a mass $2505\mev$ and a width about $32\mev$. The predicted width of our results is eight times smaller than the one reported. We should admit that there are also uncertainties in our results, seen from the results of $|T_{\rho-K_3}^{I=1/2}|^2$ on the right of Fig. \ref{fig:t5kf4rk3} and discussed later. For $|T_{\rho-K_3}^{I=3/2}|^2$ there is also resonant structure at the position about $2636\mev$, the width of which is about $171\mev$, which is not found for any $I=3/2$ $K_4$ state in the PDG. Thus, within uncertainties, we predict a new $K_4$ resonance of isospin $I=3/2$, with a mass about $2636\mev$ and a width about $171\mev$.

\subsection{Six-body interaction}

From Table \ref{tab:cases}, if we predict a $K_3$ state, seen in the Subsection \ref{fourb}, analogously to the five-body interaction, there are also two options of the cluster for the six-body interaction, the particle $f_4$ found in PDG and studied in Ref. \cite{Roca:2010tf} and the resonance $K_3$ predicted above. Under the FCA, we let a third particle of the composite resonance ($K_1$ or $f_2$) to collide with them, having (i) particle $3=K_1$, cluster $R=f_4$ ($1=f_2,\;2=f_2$), or (ii) $3=f_2$, $R=K_3$ ($1=f_2,\;2=K_1$). Since the isospin of the particles, $I_{f_2} = I_{f_4} = 0$ and $I_{K_1} = I_{K_3} = \frac{1}{2}$, we find that the total isospin of the six-body system is only $I_{total} = \frac{1}{2}$. Thus, for the first case, the $K_1$ colliding with the $f_4$ ($K-f_4$), we need calculate the amplitudes $t_1 = t_2 = t_{K_1 f_2} = T_{K_1-f_2}^{(I=1/2)}$, which have been evaluated in Subsection \ref{fourb}. For the second case, the $f_2$ colliding with the $K_3$, we evaluate the amplitudes $t_1 = t_{f_2 f_2} = T_{f_2-f_2}$ by reproducing the results of Ref. \cite{Roca:2010tf} within our formalism (we have already reproduced $T_{\rho-f_2}$ in Subsection \ref{fourb}), and $t_2 = t_{f_2 K_1} = T_{f_2-K_1}$ taking from the results of Subsection \ref{fourb}. 

\begin{figure}
\centering
\includegraphics[scale=0.6]{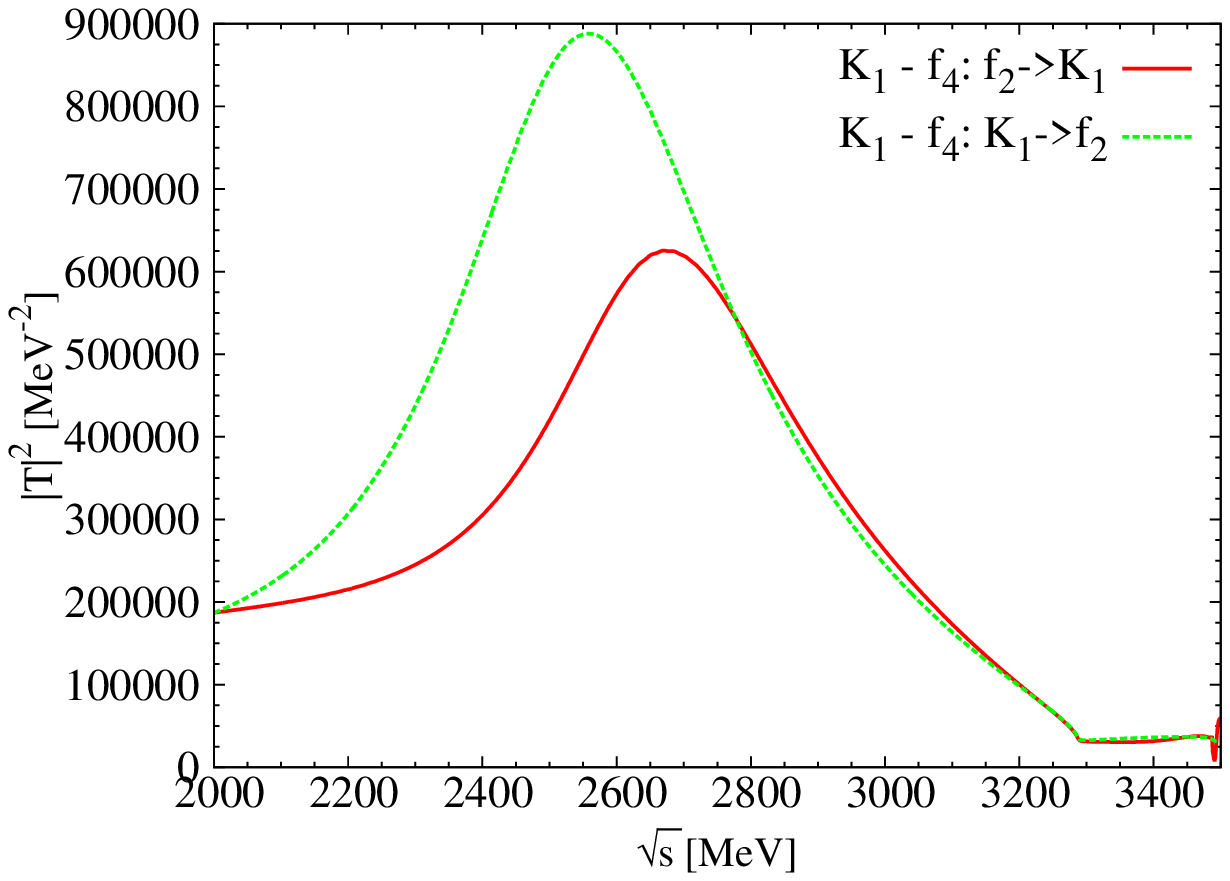}
\includegraphics[scale=0.6]{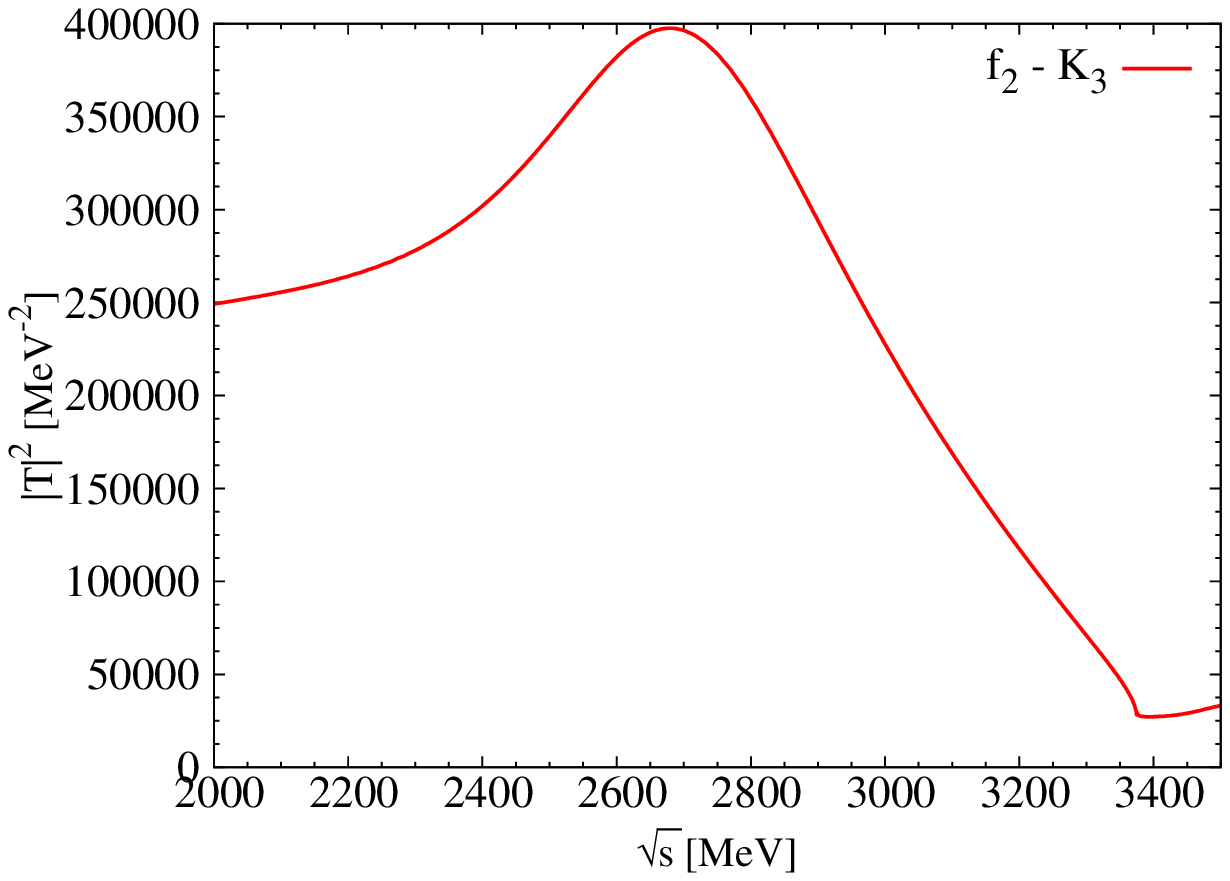}
\caption{Modulus squared of the $T_{K_1-f_4}$ (left) and $T_{f_2-K_3}$ (right) scattering amplitudes.}
\label{fig:t6k1f4f2k3}
\end{figure}

We show our results for the six-body interaction in Fig. \ref{fig:t6k1f4f2k3}. The left of Fig. \ref{fig:t6k1f4f2k3} is $|T_{K_1-f_4}^{I=1/2}|^2$. Since $K_1$ collides with the $f_4$, the amplitudes $t_1 = t_2 = t_{K_1 f_2} = T_{K_1-f_2}^{(I=1/2)}$, as shown in Subsection \ref{fourb}, the amplitude $T_{K_1-f_2}^{(I=1/2)} \neq T_{f_2-K_1}^{(I=1/2)}$, for a test we also take $t_1 = t_2 = t_{K_1 f_2} = T_{f_2-K_1}^{(I=1/2)}$ to evaluate the scattering amplitude again. By taking $t_1 = t_2 = t_{K_1 f_2} = T_{K_1-f_2}^{(I=1/2)}$, we find a peak around the energy $2558\mev$ with a large width of about $531\mev$. For the results of taking $t_1 = t_2 = t_{K_1 f_2} = T_{f_2-K_1}^{(I=1/2)}$, there also is a resonant peak in the position about $2670\mev$, the width of which is about $543\mev$. We can see that this test just gives us some uncertainties for our results. The right of Fig. \ref{fig:t6k1f4f2k3} is $|T_{f_2-K_3}^{I=1/2}|^2$, where we can observe that there is not a clear peak at the position $2681\mev$. This peak looks like the resonant structure of $f_2-K_3$ which is less stable comparing with the one of $K_1-f_4$. Since there is no $K_5$ particle found in PDG, there could be a new $K_5$ resonance which is more uncertainty from our results, with a mass of about $2558 \sim 2681\mev$ and a large width about $531 \sim 543\mev$ or more.

\subsection{Discussions}

\begin{figure}
\centering
\includegraphics[scale=0.6]{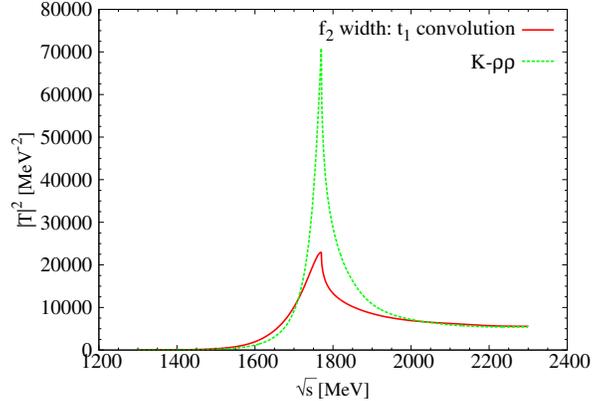}
\caption{Considering the convolution of $t_i$ in the $K-\rho\rho$ scattering amplitudes.}
\label{fig:t3krrcon}
\end{figure}

    Now we would like to make a discussion about the uncertainties of our results. As discussed in Section \ref{sectwo}, we have considered the width of the vector mesons in the evaluation of the two-body interaction amplitudes, $t_{\rho \rho}$ and $t_{\rho K}$, by taking into account the convolution of the loop function $G^I$. On the other hand, for the $K$-multi-$\rho$ system in the present work, the third particle in many cases, seen in Table. \ref{tab:cases}, is a vector meson or a resonance, and has a large width, which should be taken into account. Thus, as done in Ref. \cite{Xiao:2012dw}, we can roughly consider the width of the third particle in the propagator function $G_0$, Eq. \eqref{eq:G0s}, by replacing the term in the denominator $i \epsilon$ with $i m_3 \Gamma_3$. Or, we can do the convolution of $G_0$ function for considering more exactly the contribution of the width of the third particle as done in Ref. \cite{Bayar:2013bta}. But, as shown in the results of Refs. \cite{Xiao:2012dw,Bayar:2013bta}, the final results for the position of the generated states are not altered after considering the contribution of the width of the third particle, to which we assign some uncertainties on the width for the generated results. Furthermore, there is also certain width for the clusters, for example, the resonances of $f_2 (1270)$ and $K_1 (1270)$ in the three-body interaction, which have a big width. As done in Refs.\cite{Xiao:2012dw,Bayar:2013bta}, we also would take into account roughly the contribution of the width of the cluster by replacing the mass of the cluster $M_R$ in Eqs. \eqref{eq:formfactor} and \eqref{eq:formfactorN} with $M_R - i \Gamma_R/2$. In fact, the contribution for the width of the cluster is just a small effect on the masses and widths of the generated results (seen more discussions in Refs.\cite{Xiao:2012dw,Bayar:2013bta}).  For checking this contribution, we also can consider the convolution for the $t_i$ since Eq. \eqref{eq:si} is dependent on the mass of the cluster, seen in Fig. \ref{fig:t3krrcon} taking $K-\rho\rho$ scattering for example, where we can see that the effect of the width of the cluster is small for the position of the peak and just makes the width of the peak a bit larger.
Therefore, in our present work, we ignore all these effects in our investigation, just keeping in mind that there also some uncertainties from considering the contribution of the width of the third particle and the cluster. Besides, according to the discussions in Ref. \cite{Lensky:2005hb}, there may be some uncertain effect in the denominator term of $\tilde{t_1}~\tilde{t_2}~G_0^2$ ($\tilde{t_1}~G_0$) in Eq. \eqref{eq:new} (Eq. \eqref{eq:new2}).

We would like to make some discussions concerning the off-shell contributions in our formalism. For the two-body interactions, we follow the  works of Refs. \cite{Molina:2008jw,Geng:2006yb} for the $\rho\rho$ and $\rho K$ interactions, where the on-shell Bethe-Salpeter equation was used. A short prove of the on-shell Bethe-Salpeter equation is given in Refs. \cite{Oller:1997ti,Oset:1997it}, and a different justification is given in Ref. \cite{Oller:2000fj}, where a subtracted dispersion relation is used and one can obtain the results of the on-shell Bethe-Salpeter equation when one neglects the contribution of the left hand cut. Besides, the on-shell approximation and full off-shell effects are discussed in detail in the work of \cite{Nieves:1999bx} for the $\pi \pi$ scattering and in the works of Bonn group \cite{Borasoy:2007ku,Mai:2012wy,Mai:2014xna}. In the present work, we investigate the $K$-multi-$\rho$ interactions with the FCA to Faddeev equations, which are different from the full Faddeev equations with ChUA as done in Ref. \cite{MartinezTorres:2007sr}. One of the intrinsic differences between them is that the FCA to Faddeev equations based on the two-body on-shell amplitudes as input while the full Faddeev equations using the full off-shell amplitudes. Yet, this should not be seen as a draw back of the FCA. Indeed, in the full Faddeev equations calculations of Ref. \cite{MartinezTorres:2007sr}, the full off-shell amplitudes can be separated into two parts: an on-shell part, and a remnant off-shell part which are unphysical. In full Faddeev calculations these unphysical parts will be implicitly canceled by ad hoc three-body forces which are included in the calculations, details seen in Refs. \cite{MartinezTorres:2008gy,Khemchandani:2008rk}. This means that using the on-shell amplitudes in the Faddeev equations is a more appropriate way to deal with the three-body system than using the full-off shell amplitudes if no additional three-body forces are introduced \cite{MartinezTorres:2008kh}. The use of the on-shell amplitudes in the FCA gets a strong support from these findings. Indeed, since the on-shell two-body amplitudes are used as input in the FCA to Faddeev equations, the cancellation of the off-shell contribution can not be hold as in the full Faddeev equations calculations \cite{MartinezTorres:2007sr}. Up to one-loop, there are only two diagrams, seen (b) and (f) of Fig. \ref{fig:FCA}, while having six diagrams in the full Faddeev equations calculations. Following Ref. \cite{MartinezTorres:2007sr}, we divide the on-shell and off-shell parts for the two-body amplitudes, written $t_i (s_i) = t_i^{on} (s_i) + t_i^{off} (s_i)$, thus, using Eq. \eqref{eq:s2} we have for one-loop
\begin{equation}
T_1^{(b)} (T_2^{(f)}) \sim \int \frac{d^3q}{(2\pi)^3} F_R(\vec{q}\,) \frac{1}{{q^0}^2-\vec{q}\,^2-m_3^2+i\,\epsilon} (t_1^{on} + t_1^{off}) (t_2^{on} + t_2^{off}),
\end{equation}
where we can see that the off-shell parts of the two-body amplitudes can not be canceled since that we do not have the interaction force from the contact terms. But, one also can be see that, because of the constraints of the form factor $F_R(\vec{q}\,)$ (seen Fig. \ref{fig:formf_K1f2}), the contribution of the off-shell parts can not be infinity. The results of Ref. \cite{Bayar:2015oea} (their Fig. 6) show that the position of the peak move to higher less then 3\% when one take a $q_i$ dependent to $t_i (s_i)$. Thus, the contribution of the off-shell parts are small which can be included in our uncertainties.

\section{Conclusions}

In the present work, we investigate the many body interactions in $K$-multi-$\rho$ systems, using the formalism of the fixed center approximation to the Faddeev equations. We start from the two-body interaction of $\rho \rho$ and $\rho K$ with the combination of dynamics of the local hidden gauge Lagrangian and the coupled channel effect, to reproduce the resonances of $f_2 (1270)$ and $K_1 (1270)$ as the clusters for our formalism. We summarize our results in Table \ref{tab:summa}, where we obtain some bound states in the $K$-multi-$\rho$ interactions, as the findings of the three-body wave functions in a finite volume in Ref. \cite{Meissner:2014dea}. In the three-body $K \rho \rho$ system, we dynamically generate the $K_2 (1770)$ state in our formalism, obtained a resonant peak in the modulus squared of the scattering amplitudes around the position in $1707\mev$ with a width about $113\mev$ and explained its structure as a $\rho-K_1 (1270)$ molecular state. Continuing with the three-body Faddeev equations with the fixed center approximation, we extrapolate our formalism to the four-body interaction. We observe a new $K_3$ state, with a mass about $2079 \sim 2091\mev$ and a width about $230 \sim 249\mev$, which would be a $K_1-f2$ molecular resonance and not found in the PDG yet. For the five-body interaction, we successfully generate the $K_4 (2500)$ state in our results of $|T_{K-f_4}^{I=1/2}|^2$, with a mass $2505\mev$ and a width about $32\mev$, even though our theoretically predicted width is smaller than the one of unconfirmed experimental results of about $250\mev$. Thus, we also explain the structure of $K_4 (2500)$ as a molecular state of $K-f_4$. Besides, we find a new $K_4$ resonance of isospin $I=3/2$ in the $\rho-K_3$ interaction, with a mass about $2636\mev$ and a width about $171\mev$. Finally, analogously we predict a new $K_5$ state in the six-body interactions of the $K_1-f_4$ and $f_2-K_3$, with a mass of about $2558 \sim 2681\mev$ and a large width about $531 \sim 543\mev$ or more, with more uncertainties. We hope that in future experiments, our predicted states of $K_3 (2080)$, $K_5 (2670)$ (isospin $I=1/2$), and $K_4 (2640)$ (isospin $I=3/2$) are found.
\begin{table}[htb]
\centering
\caption{Our results of the $K$-multi-$\rho$ interactions (units: MeV).}
\label{tab:summa}
\begin{tabular}{|c|c|c|c|c|}
\hline
\hspace{0.2cm} Interaction \hspace{0.2cm} & \hspace{0.7cm} interactions \hspace{0.7cm} & \hspace{0.2cm} Results (mass, width) \hspace{0.2cm} &  \hspace{0.2cm} Exp. Res \hspace{0.2cm} &  \hspace{0.2cm} Pridictions \hspace{0.2cm}  \\
\hline
 & $K-f_2\;(\rho\rho)$ & cusp &  $-$  & $-$ \\
Three-body  & $\rho-K_1\;(\rho K) ^{I=1/2}$ & (1707, 113) &  $K_2 (1770)$  & $-$ \\
    & $\rho-K_1\;(\rho K) ^{I=3/2}$ & $-$ &  $-$  & $-$ \\
\hline
  & $f_2-K_1\;(\rho K)$ & (2079, 249)  &  $-$ & \\
\rb{Four-body}  & $K_1-f_2\;(\rho\rho)$ & (2091, 230) & $-$ & \rb{$K_3 (2080)$}\\
\hline
   & $K-f_4\;(f_2 f_2)$ & (2505, 32) &    & $-$  \\
Five-body   & $\rho-K_3\;(f_2 K_1) ^{I=1/2}$ & (2382, 409) & \rb{$K_4 (2500)$}  & $-$  \\
   & $\rho-K_3\;(f_2 K_1) ^{I=3/2}$ & (2636, 171) &  $-$ & $K_4 (2640)$\\
\hline
  & $K_1-f_4\;(f_2 f_2)$ & (2558-2670, 531-543) & $-$  & \\
\rb{Six-body}  & $f_2-K_3\;(f_2 K_1)$ & (2681, $>500$) & $-$ & \rb{$K_5 (2670)$}\\
\hline
\end{tabular}
\end{table}

\section*{Acknowledgements}

We thank M. J. Vicente Vacas for useful discussions, also appreciate E. Oset, C. Hanhart and Ulf-G. Mei{\ss}ner for careful reading the paper and useful comments.
This work is supported, in part, by NSFC and DFG through funds provided to the Sino-Germen CRC 110 “Symmetries and the Emergence of Structure in QCD (NSFC Grant No. 11261130311), NSFC (Grant Nos. 11035006 and 11165005).
This work is also partly supported by the Spanish Ministerio de Economia y Competitividad and European FEDER funds under Contract No. FIS2011-28853-C02-01 and the Generalitat Valenciana in the program Prometeo, 2009/090. We acknowledge the support of the European Community-Research Infrastructure Integrating Activity Study of Strongly Interacting Matter (Hadron Physics 3, Grant No. 283286) under the Seventh Framework Programme of the European Union.

\appendix

\section{Contribution of the box diagram and the convolution of the loop}
\label{appa}

As mention in Ref. \cite{Molina:2008jw}, we should take into account the contribution of the box diagram to the potential $V^I$. The main contribution is $\pi\pi$-box diagram for $\rho\rho$ interaction, written
\begin{equation} 
V_{box(\rho\rho)}^{(2\pi,I=0,S=2)} (s) = 8\tilde{V}^{(\pi\pi)},
\end{equation}
where $\tilde{V}^{(\pi\pi)}$ is given by
\begin{equation}
\begin{split}
\tilde{V}^{(\pi\pi)} (s) =& (\sqrt{2} g)^4 \frac{8}{15\pi^2} \int_0^{q'_{max}} dq \,\vec{q}\;^6 [10\omega^2 - (k_3^0)^2] \frac{1}{\omega^3} \Big( \frac{1}{k_1^0 + 2\omega} \Big)^2 \frac{1}{P^0 + 2\omega} \\
&\times \frac{1}{k_1^0 + \frac{\Gamma}{4} - 2\omega + i\epsilon} \frac{1}{k_1^0 - \frac{\Gamma}{4} - 2\omega + i\epsilon} \frac{1}{P^0 - 2\omega + i\epsilon} F(q)^4,
\end{split}
\end{equation}
where $\omega = \sqrt{\vec{q}\;^2+m_\pi^2}$, $\sqrt{s} = P^0 = k_1^0 + k_2^0$. For the cut off in the integration, we took a natural size, $q'_{max}=1200\mev$. Beside, $F(q)$ is a form factor for an off-shell $\pi$ in each vertex, in the case of $\pi\pi$-box, taken
\begin{equation} 
F(q) = \frac{\Lambda^2-m_\pi^2}{\Lambda^2+\vec{q}\;^2},
\end{equation}
with $\Lambda=1300\mev$. 

The element of the loop function in the matrix $G^I$ in Eq. \eqref{eq:bse} is only two meson loop, read
\begin{equation} 
G_{ii}^I (s,m_1,m_2) = i \int \frac{d^4 q}{(2\pi)^4} \frac{1}{q^2 - m_1^2 + i\epsilon} \frac{1}{(P-q)^2 - m_2^2 + i\epsilon},
\label{eq:loop}
\end{equation}
which could be regularized by the cut off, obtained
\begin{equation} 
G_{ii}^I (s,m_1,m_2) = \int_0^{q_{max}} \frac{\vec{q}\;^2 d|\vec{q}|}{(2\pi)^2} \frac{\omega_1 + \omega_2}{\omega_1 \omega_2 [(P^0)^2 - (\omega_1 + \omega_2)^2 + i\epsilon]},\label{eq:cutoff}
\end{equation}
where $\omega_i = \sqrt{\vec{q}\;^2+m_i^2}$, the centre-of-mass energy $(P^0)^2=s$, and $q_{max}$ stands for the cutoff, of which we took $q_{max}=875\mev$ for the $\rho\rho$ interaction. It is notice that we also could do the calculation of Eq. \eqref{eq:cutoff} with the analytic expressions in Refs. \cite{Oller:1998hw,Guo:2005wp}. But we need take into account the convolution due to the $\rho$ mass distribution by replacing the $G$ function as follows:
\begin{equation}
\begin{split}
\tilde{G}_{\rho\rho}(s) =& \frac{1}{N^2} \int_{(m_\rho-2\Gamma_\rho)^2}^{(m_\rho+2\Gamma_\rho)^2} d\tilde{m}_1^2 \Big( -\frac{1}{\pi} \Big) \Ima \frac{1}{\tilde{m}_1^2 - m_\rho^2 + i \tilde{m}_1 \Gamma(\tilde{m}_1)}\\
&\times \int_{(m_\rho-2\Gamma_\rho)^2}^{(m_\rho+2\Gamma_\rho)^2} d\tilde{m}_2^2 \Big( -\frac{1}{\pi} \Big) \Ima \frac{1}{\tilde{m}_2^2 - m_\rho^2 + i \tilde{m}_2 \Gamma(\tilde{m}_2)} G_{\rho\rho}(s,\tilde{m}_1^2,\tilde{m}_2^2),
\end{split}
\end{equation}
with 
\begin{align}
&N= \int_{(m_\rho-2\Gamma_\rho)^2}^{(m_\rho+2\Gamma_\rho)^2} d\tilde{m}_1^2 \Big( -\frac{1}{\pi} \Big) \Ima \frac{1}{\tilde{m}_1^2 - m_\rho^2 + i \tilde{m}_1 \Gamma(\tilde{m}_1)},\\
&\Gamma(\tilde{m}_1)= \Gamma_\rho \Big( \frac{\tilde{m}_1^2 - 4m_\pi^2}{m_\rho^2 - 4m_\pi^2} \Big)^{3/2},
\end{align}
where $\Gamma_\rho = 146.2\mev$ and $G_{\rho\rho} (s,\tilde{m}_1^2,\tilde{m}_2^2)$, is given by Eq. \eqref{eq:cutoff}.

\section{Coefficients of the potential}
\label{appb}

The coefficients $C_{ij}$ in Eq.~\eqref{eq:V_VVPP} for the $I=1/2$ sector are given and tabulated in Table \ref{table:rhoK1/2}.
In the $I=3/2$ sector, the relevant channels are $\rho K$ and $K^* \pi$ and the coefficients are given in Table~\ref{table:rhoK3/2}.
  
   \begin{table}
   \caption{ Coefficients $C_{ij}$ in eq.~\eqref{eq:V_VVPP} in the $I=1/2$ sector.}
   \begin{center}
   \begin{tabular}{c|ccccc}
   \hline
   \hline
                & $\phi K$ & $\omega K$ & $\rho K$ & $K^* \eta$ & $K^* \pi$ \\
   \hline
   $\phi K$     & 0 & 0 & 0 & $-\sqrt{\frac{3}{2}}$ & $-\sqrt{\frac{3}{2}}$ \\
   $\omega K$   & 0 & 0 & 0 & $\frac{\sqrt{3}}{2}$ & $\frac{\sqrt{3}}{2}$ \\
   $\rho K$     & 0 & 0 & $-2$ & $-\frac{3}{2}$ & $ \frac{1}{2}$ \\
   $K^* \eta$   & $-\sqrt{\frac{3}{2}}$ & $\frac{\sqrt{3}}{2}$ & $-\frac{3}{2}$ & 0 & 0\\
   $K^* \pi$    & $-\sqrt{\frac{3}{2}}$ & $\frac{\sqrt{3}}{2}$ & $\frac{1}{2}$ & 0 & $-2$ \\
   \hline
   \hline
   \end{tabular}
   \label{table:rhoK1/2}
   \end{center}
   \end{table}

   \begin{table}
   \caption{ Coefficients $C_{ij}$ in eq.~\eqref{eq:V_VVPP} in the $I=3/2$ sector.}
   \begin{center}
   \begin{tabular}{c|cc}
   \hline
   \hline
              & $\rho K$ & $K^* \pi $\\
   \hline
   $\rho K$   & 1 & 1 \\
   $K^* \pi$  & 1 & 1 \\
   \hline
   \hline
   \end{tabular}
   \label{table:rhoK3/2}
   \end{center}
   \end{table}

\section{Convolution of $G^I$ loop function}
\label{appc}

Taking the on-shell factorization, the loop function for two mesons is expressed as a function of the energy $s$, seen Eq.~\eqref{eq:loop}.
To remove the ultra violet divergence of the loop function, we follow the dimensional regularization scheme
  \begin{eqnarray}
  G_{ll}(s,M_l,m_l)
  &=&
  \frac{1}{16\pi^2}
  \left\{
  a(\mu) + {\rm ln} \frac{M_l^2}{\mu^2} + \frac{m_l^2 - M_l^2 + s}{2s} {\rm ln}\frac{m_l^2}{M_l^2}
  \right. \nonumber \\
  &&
  + \frac{q_l}{\sqrt{s}}
  \left[
  {\rm ln} (s - (M_l^2 - m_l^2) + 2q_l \sqrt{s})
  \right. \nonumber \\
  &&
  + {\rm ln} (s + (M_l^2 - m_l^2) + 2q_l \sqrt{s})
  \nonumber \\
  &&
  -{\rm ln} (-s + (M_l^2 - m_l^2) + 2q_l \sqrt{s})
  \nonumber \\
  &&
  \left.
  \left.
  -{\rm ln} (-s - (M_l^2 - m_l^2) + 2q_l \sqrt{s})
  \right]
  \right\}
  \label{eq:G_rhoK},
  \end{eqnarray}
with a momentum $q_l$ determined at the center of mass frame
  \begin{eqnarray}
  q_l
  =
  \frac{ \sqrt{ [s- (M_l-m_l)^2][s - (M_l+m_l)^2)] }}{ 2\sqrt{s} },
  \end{eqnarray}
where $\mu$ is a scale parameter in this scheme.
The finite part of the loop function is stable against changes of $\mu$ due to the subtraction constant $a(\mu)$ which absorbs the changes of $\mu$, where we take the following parameter set chosen to reproduce $K_1 (1270)$ in Ref.~\cite{Geng:2006yb}
  \begin{eqnarray}
  \mu=900~{\rm MeV},~ a(\mu)=-1.85,~f=115~{\rm MeV}.
  \end{eqnarray}

But, considering a finite width of the vector mesons in the loop function,
as done in Ref.~\cite{Geng:2006yb}, the effect of the propagation of unstable particles is taken into account in terms of the Lehmann representation,
which is done by the dispersion relation with its imaginary part
  \begin{eqnarray}
  D(s)
  =
  \int_{ s_{ {\rm th} }}^{\infty}
  ds_V
  \left( - \frac{1}{\pi} \right)
  \frac{{\rm Im} D(s_V)}{s - s_V + i\epsilon }
  \label{eq:disp},
  \end{eqnarray}
where $s_{{\rm th}}$ stands for the square of the threshold energy.
Now the spectral function is taken as
  \begin{eqnarray}
  {\rm Im}D(s_V)
  =
  {\rm Im} \left\{ \frac{1}{s_V - M_V^2 + iM_V \Gamma_V} \right\}
  \label{eq:im_disp},
  \end{eqnarray}
where the width $\Gamma_V$ is assumed to be a constant physical value.
Substituting Eqs.~\eqref{eq:disp} and \eqref{eq:im_disp} into the original loop function Eq.~\eqref{eq:G_rhoK}, we have
  \begin{eqnarray}
  \tilde{G}_{ll}(s,M_l,m_l)
  &=&
  \frac{1}{C_l}
  \int_{(M_l - 2 \Gamma_l)^2}^{(M_l + 2\Gamma_l)^2}
  ds_V G_{ll} (s,\sqrt{s_V},m_l) \nonumber \\
  &\times&
  \left( - \frac{1}{\pi} \right)
  {\rm Im} \left\{ \frac{1}{s_V - M_l^2 + iM_l \Gamma_l} \right\} ,
  \end{eqnarray}
where $G_{ll}$ is given by Eq.~\eqref{eq:G_rhoK}, and the normalization for the $l$th component
  \begin{eqnarray}
  C_l
  &=&
  \int_{(M_l - 2 \Gamma_l)^2}^{(M_l + 2\Gamma_l)^2}
  ds_V \times
  \left( - \frac{1}{\pi} \right)
  {\rm Im} \left\{ \frac{1}{s_V - M_l^2 + iM_l \Gamma_l} \right\} , 
  \end{eqnarray}
with $m_l$, $M_l$, $\Gamma_l$, the mass of the pseudoscalar meson, mass of the vector and width of the vector respectively.
Replacing $G_{ll}$ by $\tilde{G}_{ll}$ in Eq.~\eqref{eq:G_rhoK}, we include the width effect of vector mesons. In the present case, we only do the convolution for the $\rho$ and $K^*$.


\begin{thebibliography}{99}
\bibitem{Feynman:1964fk} 
  R.~P.~Feynman, M.~Gell-Mann and G.~Zweig,
  Phys.\ Rev.\ Lett.\  {\bf 13}, 678 (1964).

\bibitem{GellMann:1964nj} 
  M.~Gell-Mann,
  Phys.\ Lett.\  {\bf 8}, 214 (1964).

\bibitem{Olsen:2014mea} 
  S.~L.~Olsen,
  Hyperfine Interact.\  {\bf 229}, no. 1-3, 7 (2014).

\bibitem{Choi:2014iwa} 
  S.~Choi,
  Int.\ J.\ Mod.\ Phys.\ Conf.\ Ser.\  {\bf 31}, 1460293 (2014).

\bibitem{Gasser:1984gg} 
  J.~Gasser and H.~Leutwyler,
  Nucl.\ Phys.\ B {\bf 250}, 465 (1985).

\bibitem{Meissner:1993ah} 
  U.~-G.~Mei{\ss}ner,
  Rept.\ Prog.\ Phys.\  {\bf 56}, 903 (1993).

\bibitem{Bernard:1995dp}
 V.~Bernard, N.~Kaiser and U.-G.~Mei{\ss}ner,
 Int.\ J.\ Mod.\ Phys.\ E {\bf 4}, 193 (1995).

\bibitem{Pich:1995bw} 
  A.~Pich,
  Rept.\ Prog.\ Phys.\  {\bf 58}, 563 (1995).

\bibitem{Ecker:1994gg} 
  G.~Ecker,
  Prog.\ Part.\ Nucl.\ Phys.\  {\bf 35}, 1 (1995).

\bibitem{Scherer:2002tk} 
  S.~Scherer,
  Adv.\ Nucl.\ Phys.\  {\bf 27}, 277 (2003).

\bibitem{Bernard:2007zu}
 V.~Bernard,
 Prog.\ Part.\ Nucl.\ Phys.\  {\bf 60}, 82 (2008).

\bibitem{Politzer:1988bs}
  H.~D.~Politzer and M.~B.~Wise,
  Phys.\ Lett.\ B {\bf 208} (1988) 504.

\bibitem{Georgi:1990um}
  H.~Georgi,
  Phys.\ Lett.\ B {\bf 240}, 447 (1990).

\bibitem{Epelbaum:2008ga}
  E.~Epelbaum, H.~-W.~Hammer and U.~-G.~Meissner,
  Rev.\ Mod.\ Phys.\  {\bf 81}, 1773 (2009).

\bibitem{Kogut:1982ds}
  J.~B.~Kogut,
  Rev.\ Mod.\ Phys.\  {\bf 55}, 775 (1983).

\bibitem{Luscher:1996sc}
  M.~Luscher, S.~Sint, R.~Sommer and P.~Weisz,
  Nucl.\ Phys.\ B {\bf 478}, 365 (1996).

\bibitem{Luscher:1996ug}
  M.~Luscher, S.~Sint, R.~Sommer, P.~Weisz and U.~Wolff,
  Nucl.\ Phys.\ B {\bf 491}, 323 (1997).

\bibitem{Shifman:1978bx}
  M.~A.~Shifman, A.~I.~Vainshtein and V.~I.~Zakharov,
  Nucl.\ Phys.\ B {\bf 147}, 385 (1979).

\bibitem{Reinders:1984sr}
  L.~J.~Reinders, H.~Rubinstein and S.~Yazaki,
  Phys.\ Rept.\  {\bf 127}, 1 (1985).

\bibitem{Dias:2012ek}
  J.~M.~Dias, R.~M.~Albuquerque, M.~Nielsen and C.~M.~Zanetti,
  Phys.\ Rev.\ D {\bf 86}, 116012 (2012).

\bibitem{Zhou:2014ytp} 
  D.~Zhou, E.~L.~Cui, H.~X.~Chen, L.~S.~Geng, X.~Liu and S.~L.~Zhu,
  Phys.\ Rev.\ D {\bf 90}, 114035 (2014).

\bibitem{Roberts:1994dr}
  C.~D.~Roberts and A.~G.~Williams,
  Prog.\ Part.\ Nucl.\ Phys.\  {\bf 33}, 477 (1994).

\bibitem{Maris:2003vk}
  P.~Maris and C.~D.~Roberts,
  Int.\ J.\ Mod.\ Phys.\ E {\bf 12}, 297 (2003).

\bibitem{Fischer:2006ub}
  C.~S.~Fischer,
  J.\ Phys.\ G {\bf 32}, R253 (2006).

\bibitem{Manohar:1983md}
  A.~Manohar and H.~Georgi,
  Nucl.\ Phys.\ B {\bf 234}, 189 (1984).

\bibitem{Zhang:1997ny}
  Z.~Y.~Zhang, Y.~W.~Yu, P.~N.~Shen, L.~R.~Dai, A.~Faessler and U.~Straub,
  Nucl.\ Phys.\ A {\bf 625}, 59 (1997).

\bibitem{Fontoura:2012mz} 
  C.~E.~Fontoura, G.~Krein and V.~E.~Vizcarra,
  Phys.\ Rev.\ C {\bf 87}, 025206 (2013).

\bibitem{Oller:1997ti} 
  J.~A.~Oller and E.~Oset,
  Nucl.\ Phys.\ A {\bf 620}, 438 (1997)
  [Erratum-ibid.\ A {\bf 652}, 407 (1999)].

\bibitem{Oset:1997it} 
  E.~Oset and A.~Ramos,
  Nucl.\ Phys.\ A {\bf 635}, 99 (1998).

\bibitem{Oller:1997ng} 
  J.~A.~Oller, E.~Oset and J.~R.~Pelaez,
  Phys.\ Rev.\ Lett.\  {\bf 80}, 3452 (1998).

\bibitem{Oller:1998hw} 
  J.~A.~Oller, E.~Oset and J.~R.~Pelaez,
  Phys.\ Rev.\ D {\bf 59}, 074001 (1999)
  [Erratum-ibid.\ D {\bf 60}, 099906 (1999)]
  [Erratum-ibid.\ D {\bf 75}, 099903 (2007)].

\bibitem{Oller:1998zr} 
  J.~A.~Oller and E.~Oset,
  Phys.\ Rev.\ D {\bf 60}, 074023 (1999).

\bibitem{Oller:2000fj}
 J.~A.~Oller and U.-G.~Mei{\ss}ner,
 Phys.\ Lett.\ B {\bf 500}, 263 (2001).

\bibitem{Fujita:1957zz}
  J.~Fujita and H.~Miyazawa,
  Prog.\ Theor.\ Phys.\  {\bf 17}, 360 (1957).

\bibitem{Faddeev:1960su}
  L.~D.~Faddeev,
  Sov.\ Phys.\ JETP {\bf 12}, 1014 (1961)
  [Zh.\ Eksp.\ Teor.\ Fiz.\  {\bf 39}, 1459 (1960)].

\bibitem{Glockle:1986zz}
  W.~Glockle, T.~S.~H.~Lee and F.~Coester,
  Phys.\ Rev.\ C {\bf 33}, 709 (1986).

\bibitem{Weinberg:1992yk}
  S.~Weinberg,
  Phys.\ Lett.\ B {\bf 295}, 114 (1992).

\bibitem{Richard:1992uk}
  J.~M.~Richard,
  Phys.\ Rept.\  {\bf 212}, 1 (1992).

\bibitem{MartinezTorres:2007sr} 
  A.~Martinez Torres, K.~P.~Khemchandani and E.~Oset,
  Phys.\ Rev.\ C {\bf 77}, 042203 (2008).

\bibitem{MartinezTorres:2012jr}
  A.~Martinez Torres, K.~P.~Khemchandani, M.~Nielsen and F.~S.~Navarra,
  Phys.\ Rev.\ D {\bf 87}, 034025 (2013).

\bibitem{Roca:2010tf}
  L.~Roca and E.~Oset,
  Phys.\ Rev.\ D {\bf 82}, 054013 (2010).

\bibitem{Toker:1981zh}
  G.~Toker, A.~Gal and J.~M.~Eisenberg,
  Nucl.\ Phys.\ A {\bf 362}, 405 (1981).

\bibitem{Barrett:1999cw}
  R.~C.~Barrett and A.~Deloff,
  Phys.\ Rev.\ C {\bf 60}, 025201 (1999).

\bibitem{Deloff:1999gc}
  A.~Deloff,
  Phys.\ Rev.\ C {\bf 61}, 024004 (2000).

\bibitem{Kamalov:2000iy}
  S.~S.~Kamalov, E.~Oset and A.~Ramos,
  Nucl.\ Phys.\ A {\bf 690}, 494 (2001).

\bibitem{Gal:2006cw}
  A.~Gal,
  Int.\ J.\ Mod.\ Phys.\ A {\bf 22}, 226 (2007).

\bibitem{YamagataSekihara:2010qk} 
  J.~Yamagata-Sekihara, L.~Roca and E.~Oset,
  Phys.\ Rev.\ D {\bf 82}, 094017 (2010)
  [Erratum-ibid.\ D {\bf 85}, 119905 (2012)].

\bibitem{Xiao:2012dw} 
  C.~W.~Xiao, M.~Bayar and E.~Oset,
  Phys.\ Rev.\ D {\bf 86}, 094019 (2012).

\bibitem{Aaij:2015sqa} 
  R.~Aaij {\it et al.}  [LHCb Collaboration],
  arXiv:1505.01710 [hep-ex].

\bibitem{Bayar:2011qj} 
  M.~Bayar, J.~Yamagata-Sekihara and E.~Oset,
  Phys.\ Rev.\ C {\bf 84}, 015209 (2011).

\bibitem{Bayar:2012rk}
  M.~Bayar and E.~Oset,
  Nucl.\ Phys.\ A {\bf 883}, 57 (2012).

\bibitem{MartinezTorres:2008kh}
  A.~Martinez Torres, K.~P.~Khemchandani and E.~Oset,
  Phys.\ Rev.\ C {\bf 79}, 065207 (2009).

\bibitem{Jido:2008kp}
  D.~Jido and Y.~Kanada-En'yo,
  Phys.\ Rev.\ C {\bf 78}, 035203 (2008).

\bibitem{Shevchenko:2014uva} 
  N.~V.~Shevchenko and J.~Revai,
  Phys.\ Rev.\ C {\bf 90}, 034003 (2014).

\bibitem{Revai:2014twa} 
  J.~Revai and N.~V.~Shevchenko,
  Phys.\ Rev.\ C {\bf 90}, 034004 (2014).

\bibitem{Miyagawa:2012xz} 
  K.~Miyagawa and J.~Haidenbauer,
  Phys.\ Rev.\ C {\bf 85}, 065201 (2012).

\bibitem{Jido:2012cy} 
  D.~Jido, E.~Oset and T.~Sekihara,
  Eur.\ Phys.\ J.\ A {\bf 49}, 95 (2013).

\bibitem{Mai:2014uma}
 M.~Mai, V.~Baru, E.~Epelbaum and A.~Rusetsky,
 arXiv:1411.4881 [nucl-th].

\bibitem{Bayar:2012dd} 
  M.~Bayar, C.~W.~Xiao, T.~Hyodo, A.~Dote, M.~Oka and E.~Oset,
  Phys.\ Rev.\ C {\bf 86}, 044004 (2012).

\bibitem{pdg2014}
   K.~A.~Olive {\it et al.} (Particle Data Group), Chin. Phys. C, {\bf 38}, 090001 (2014).

\bibitem{Bando:1984ej} 
  M.~Bando, T.~Kugo, S.~Uehara, K.~Yamawaki and T.~Yanagida,
  Phys.\ Rev.\ Lett.\  {\bf 54}, 1215 (1985).

\bibitem{Bando:1987br} 
  M.~Bando, T.~Kugo and K.~Yamawaki,
  Phys.\ Rept.\  {\bf 164}, 217 (1988).

\bibitem{Meissner:1987ge} 
  U.~-G.~Mei{\ss}ner,
  Phys.\ Rept.\  {\bf 161}, 213 (1988).

\bibitem{Harada:2003jx}
  M.~Harada and K.~Yamawaki,
  Phys.\ Rept.\  {\bf 381}, 1 (2003).

\bibitem{Molina:2008jw} 
  R.~Molina, D.~Nicmorus and E.~Oset,
  Phys.\ Rev.\ D {\bf 78}, 114018 (2008).

\bibitem{Roca:2005nm} 
  L.~Roca, E.~Oset and J.~Singh,
  Phys.\ Rev.\ D {\bf 72}, 014002 (2005).

\bibitem{Geng:2006yb} 
  L.~S.~Geng, E.~Oset, L.~Roca and J.~A.~Oller,
  Phys.\ Rev.\ D {\bf 75}, 014017 (2007).

\bibitem{Gamermann:2009uq}
  D.~Gamermann, J.~Nieves, E.~Oset and E.~Ruiz Arriola,
  Phys.\ Rev.\ D {\bf 81}, 014029 (2010).

\bibitem{YamagataSekihara:2010pj}
 J.~Yamagata-Sekihara, J.~Nieves, E.~Oset,
 Phys.\ Rev.\ D {\bf 83}, 014003 (2011).

\bibitem{Aceti:2012dd} 
  F.~Aceti and E.~Oset,
  Phys.\ Rev.\ D {\bf 86}, 014012 (2012).

\bibitem{Polchinski:1983gv} 
  J.~Polchinski,
  Nucl.\ Phys.\ B {\bf 231}, 269 (1984).

\bibitem{Varin:2006de} 
  T.~Varin, D.~Davesne, M.~Oertel and M.~Urban,
  Nucl.\ Phys.\ A {\bf 791}, 422 (2007)
  [hep-ph/0611220].

\bibitem{Xie:2010ig}
  J.~-J.~Xie, A.~Martinez Torres and E.~Oset,
  Phys.\ Rev.\ C {\bf 83}, 065207 (2011).

\bibitem{mandl}
F.~Mandl and G.~Shaw,
{\it Quantum Field Theory~(Wiley-Interscience, New York, 1984)}.

\bibitem{MartinezTorres:2010ax} 
  A.~Martinez Torres, E.~J.~Garzon, E.~Oset and L.~R.~Dai,
  Phys.\ Rev.\ D {\bf 83}, 116002 (2011).

\bibitem{Wu:2010jy} 
  J.~J.~Wu, R.~Molina, E.~Oset and B.~S.~Zou,
  Phys.\ Rev.\ Lett.\  {\bf 105}, 232001 (2010).

\bibitem{Wu:2010vk} 
  J.~J.~Wu, R.~Molina, E.~Oset and B.~S.~Zou,
  Phys.\ Rev.\ C {\bf 84}, 015202 (2011).

\bibitem{Xiao:2011rc} 
  C.~W.~Xiao, M.~Bayar and E.~Oset,
  Phys.\ Rev.\ D {\bf 84}, 034037 (2011).

\bibitem{Oset:2001cn}
  E.~Oset, A.~Ramos and C.~Bennhold,
  Phys.\ Lett.\  B {\bf 527}, 99 (2002)
  [Erratum-ibid.\  B {\bf 530}, 260 (2002)].

\bibitem{Guo:2005wp}
  F.~K.~Guo, R.~G.~Ping, P.~N.~Shen, H.~C.~Chiang and B.~S.~Zou,
  Nucl.\ Phys.\  A {\bf 773}, 78 (2006).

\bibitem{Liang:2013yta} 
  W.~Liang, C.~W.~Xiao and E.~Oset,
  Phys.\ Rev.\ D {\bf 88}, no. 11, 114024 (2013).

\bibitem{Guo:2014iya} 
  F.~K.~Guo, C.~Hanhart, Q.~Wang and Q.~Zhao,
  arXiv:1411.5584 [hep-ph].

\bibitem{Hyodo:2014bda} 
  T.~Hyodo,
  Phys.\ Rev.\ C {\bf 90}, no. 5, 055208 (2014).

\bibitem{Chung:1974wb} 
  S.~U.~Chung, R.~L.~Eisner, S.~D.~Protopopescu, N.~P.~Samios and R.~C.~Strand,
  Phys.\ Lett.\ B {\bf 51}, 413 (1974).

\bibitem{AguilarBenitez:1970up} 
  M.~Aguilar-Benitez, V.~E.~Barnes, D.~Bassano, S.~U.~Chung, R.~L.~Eisner, E.~Flaminio, J.~B.~Kinson and R.~B.~Palmer {\it et al.},
  Phys.\ Rev.\ Lett.\  {\bf 25}, 54 (1970).

\bibitem{Blieden:1972aj} 
  H.~R.~Blieden, G.~Finocchiaro, J.~Kirz, C.~Nef, R.~Thun, D.~Bowen, D.~Earles and W.~Faissler {\it et al.},
  Phys.\ Lett.\ B {\bf 39}, 668 (1972).

\bibitem{Tikhomirov:2003gg} 
  G.~D.~Tikhomirov, I.~A.~Erofeev, O.~N.~Erofeeva and V.~N.~Luzin,
  Phys.\ Atom.\ Nucl.\  {\bf 66}, 828 (2003)
  [Yad.\ Fiz.\  {\bf 66}, 860 (2003)].

\bibitem{Cleland:1980ya} 
  W.~E.~Cleland, A.~Delfosse, P.~A.~Dorsaz, J.~L.~Gloor, M.~N.~Kienzle-Focacci, G.~Mancarella, A.~D.~Martin and M.~Martin {\it et al.},
  Nucl.\ Phys.\ B {\bf 184}, 1 (1981).

\bibitem{Armstrong:1983gt} 
  T.~Armstrong {\it et al.}  [Bari-Birmingham-CERN-Milan-Paris-Pavia Collaboration],
  Nucl.\ Phys.\ B {\bf 227}, 365 (1983).

\bibitem{Bayar:2013bta} 
  M.~Bayar, W.~H.~Liang, T.~Uchino and C.~W.~Xiao,
  Eur.\ Phys.\ J.\ A {\bf 50}, 67 (2014).

\bibitem{Lensky:2005hb} 
  V.~Lensky, V.~Baru, J.~Haidenbauer, C.~Hanhart, A.~E.~Kudryavtsev and U.-G.~Mei{\ss}ner,
  Eur.\ Phys.\ J.\ A {\bf 26}, 107 (2005).

\bibitem{Nieves:1999bx} 
  J.~Nieves and E.~Ruiz Arriola,
  Nucl.\ Phys.\ A {\bf 679}, 57 (2000).

\bibitem{Borasoy:2007ku} 
  B.~Borasoy, P.~C.~Bruns, U.~-G.~Mei{\ss}ner and R.~Nissler,
  Eur.\ Phys.\ J.\ A {\bf 34}, 161 (2007).

\bibitem{Mai:2012wy} 
  M.~Mai, P.~C.~Bruns and U.~-G.~Mei{\ss}ner,
  Phys.\ Rev.\ D {\bf 86}, 094033 (2012).

\bibitem{Mai:2014xna} 
  M.~Mai and U.~-G.~Mei{\ss}ner,
  Eur.\ Phys.\ J.\ A {\bf 51}, 30 (2015).

\bibitem{MartinezTorres:2008gy} 
  A.~Martinez Torres, K.~P.~Khemchandani, L.~S.~Geng, M.~Napsuciale and E.~Oset,
  Phys.\ Rev.\ D {\bf 78}, 074031 (2008).

\bibitem{Khemchandani:2008rk} 
  K.~P.~Khemchandani, A.~Martinez Torres and E.~Oset,
  Eur.\ Phys.\ J.\ A {\bf 37}, 233 (2008).

\bibitem{Meissner:2014dea} 
  U.~-G.~Mei{\ss}ner, G.~R\'ios and A.~Rusetsky,
  Phys.\ Rev.\ Lett.\  {\bf 114}, 091602 (2015).

\bibitem{Bayar:2015oea} 
  M.~Bayar, X.~L.~Ren and E.~Oset,
  Eur.\ Phys.\ J.\ A {\bf 51}, no. 5, 61 (2015)
  [arXiv:1501.02962 [hep-ph]].

\end{thebibliography}
\end{document}